\documentclass{aa}
\usepackage{subfig,graphicx}
\usepackage[varg]{txfonts}
\usepackage{color}
\usepackage{longtable}
\usepackage{amsmath}
\usepackage{natbib,twoopt}
\usepackage[breaklinks=true]{hyperref} 
\bibpunct{(}{)}{;}{a}{}{,}             
\makeatletter
  \newcommandtwoopt{\citeads}[3][][]{\href{http://adsabs.harvard.edu/abs/#3}%
    {\def\hyper@linkstart##1##2{}%
     \let\hyper@linkend\@empty\citealp[#1][#2]{#3}}}
  \newcommandtwoopt{\citepads}[3][][]{%
  \nonstopmode%
   \href{http://adsabs.harvard.edu/abs/#3}%
    {\def\hyper@linkstart##1##2{}%
     \let\hyper@linkend\@empty\citep[#1][#2]{#3}}%
   \errorstopmode}
 \newcommandtwoopt{\citetads}[3][][]{%
 \nonstopmode%
  \href{http://adsabs.harvard.edu/abs/#3}%
  {\def\hyper@linkstart##1##2{}%
  \let\hyper@linkend\@empty\citet[#1][#2]{#3}}%
  \errorstopmode}
  \newcommandtwoopt{\citeyearads}[3][][]%
    {\href{http://adsabs.harvard.edu/abs/#3}
    {\def\hyper@linkstart##1##2{}%
     \let\hyper@linkend\@empty\citeyear[#1][#2]{#3}}}
\makeatother

\begin{document}

\title{Interplay between pulsations and mass loss in the blue supergiant
 55\,Cygnus = HD\,198\,478\thanks{Based on observations taken with the Perek
2m telescope at Ond\v{r}ejov Observatory, Czech Republic, and the Poznan
Spectroscopic Telescope 2 at the Winer Observatory in Arizona, USA}}

\author{M.~Kraus\inst{1}, M.~Haucke\inst{2}, L.~S.~Cidale\inst{2,3},
R.~O.~J.~Venero\inst{2,3}, D.~H.~Nickeler\inst{1}, P.~N\'{e}meth\inst{4},
E.~Niemczura\inst{5}, S.~Tomi\'{c}\inst{1,6}, \\
A.~Aret\inst{7},
J.~Kub\'{a}t\inst{1}, B.~Kub\'{a}tov\'{a}\inst{1,8}, M.~E.~Oksala\inst{9,1},
M.~Cur\'{e}\inst{10},
K.~Kami\'{n}ski\inst{11}, W.~Dimitrov\inst{11}, M.~Fagas\inst{11},\\
\and M.~Poli\'{n}ska\inst{11}
}

\institute{Astronomick\'y \'ustav, Akademie v\v{e}d \v{C}esk\'e republiky,
Fri\v{c}ova 298, 25165 Ond\v{r}ejov, Czech Republic\\
\email{michaela.kraus@asu.cas.cz}
\and
Departamento de Espectroscop\'ia Estelar, Facultad de Ciencias
Astron\'omicas y Geof\'isicas, Universidad Nacional de La Plata (UNLP), Paseo del
Bosque s/n, B1900FWA, La Plata, Argentina
\and
Instituto de Astrof\'isica de La Plata, CCT La Plata, CONICET-UNLP,
Paseo del Bosque s/n, B1900FWA, La Plata, Argentina
\and
Dr. Remeis Sternwarte, Universit\"{a}t Erlangen-N\"{u}rnberg, Sternwartstr.
7, 96049 Bamberg, Germany
\and
Astronomical Institute, Wroc\l aw University, Kopernika 11, 51-622
Wroc\l aw, Poland
\and
Matematicko fyzik\'{a}ln\'{i} fakulta, Univerzita Karlova, Praha, Czech Republic
\and
Tartu Observatory, T\~oravere, 61602 Tartumaa, Estonia
\and
Matemati\v{c}ki Institut SANU, Kneza Mihaila 36, 11001 Beograd, Serbia
\and
LESIA, Observatoire de Paris, CNRS UMR 8109, UPMC, Universit\'{e} Paris Diderot, 5 place Jules Janssen, 92195 Meudon, France
\and
 Instituto de F\'{i}sica y Astronom\'{i}a, Facultad de Ciencias, Universidad
de Valpara\'{i}so, Av. Gran Breta\~na 1111,
5030 Casilla Valpara\'{i}so, Chile
\and
Astronomical Observatory Institute, Faculty of Physics, A. Mickiewicz
University, S\l oneczna 36, 60-286 Pozna\'{n}, Poland
}

\date{Received; accepted}

\authorrunning{Kraus et al.}
\titlerunning{Interplay between pulsations and mass loss in 55\,Cyg}

\abstract {Blue supergiant stars are known to display photometric and 
spectroscopic variability that is suggested to be linked to stellar 
pulsations. Pulsational activity in massive stars strongly depends on 
the star's evolutionary stage and is assumed to be connected with mass-loss 
episodes, the appearance of macroturbulent line broadening, and the formation 
of clumps in the wind.}
{To investigate a possible interplay between pulsations and
mass-loss, we carried out an observational campaign of the supergiant 55\,Cyg
over a period of five years to search for photospheric activity and cyclic mass-loss variability in the stellar wind.}
{We modeled the H, \ion{He}{i}, \ion{Si}{ii,} and \ion{Si}{iii} lines using the nonlocal thermal equilibrium atmosphere code FASTWIND and derived the photospheric and wind parameters. In addition, we searched for variability in the intensity and radial velocity of photospheric lines and performed a moment analysis of the line profiles to derive frequencies and amplitudes of the variations.}
{The H$\alpha$ line varies with time in both intensity and shape, displaying various types of profiles: P\,Cygni, pure emission, almost complete absence, and double or multiple peaked. The star undergoes episodes of variable mass-loss rates that change by a factor of 1.7--2 on different timescales. We also observe changes in the ionization rate of \ion{Si}{ii} and 
determine a multiperiodic oscillation in the \ion{He}{i} absorption lines, with periods ranging from a few hours to 22.5 days.} 
{We interpret the photospheric line variations in terms of oscillations in p-, 
g-, and strange modes. We suggest that these pulsations can lead to phases of enhanced mass loss. Furthermore, they can mislead the determination of the stellar rotation. We classify the star as a post-red supergiant, belonging to the group of $\alpha$ Cyg variables.}

\keywords{Stars: early-type -- Stars: supergiants -- Stars: winds, outflows -- Stars: mass-loss -- Stars: activity -- Stars: individual: \object{55\,Cyg}}

\maketitle

\section{Introduction}
\label{sec:intro}

Blue supergiants (BSGs) are evolved, luminous objects that 
often display strong photometric and
spectroscopic variability \citepads[e.g.,][]{1973ApJ...186..909R, Kaufer1997,
Waelkens1998, Mathias2001, Kaufer2006, Lefever2007, Clark2010}. 
Long-term space-based photometry has linked this variability to stellar 
pulsations, establishing a new instability domain in the Hertzsprung-Russell 
(HR) diagram region populated by BSGs \citepads{Saio2006}.
The presence of pulsations in BSGs requires that excited modes are reflected 
either at an intermediate convection zone connected to the hydrogen-burning 
shell \citepads{Saio2006, 2009MNRAS.396.1833G}, or at the top of the chemical 
composition gradient region surrounding the radiative He core 
\citepads{2013MNRAS.432.3153D}.

While BSGs in the vicinity of the main sequence are found to pulsate in both
pressure (p-modes with periods of hours to about one day) and gravity modes
(g-modes with periods of 2-10\,days), objects slightly more evolved to the red 
are predicted to pulsate in pure g-modes  \citepads{Saio2006}. However, 
\citetads{Kraus2012}
recently discovered a periodic variability of 1.6\,hours in the late-type BSG
star \object{HD\,202\,850}. 
Existing pulsation models do not predict such short periods;
this emphasizes the 
current deficiency in our knowledge of the pulsation activity in BSGs.

As massive stars can cross the BSG domain more than once,
the pulsational activity of these stars can drastically 
change between their red- and blueward evolution 
\citepads{Saio2013}. 
When compared with their less-evolved counterparts, BSGs on a blue loop, or 
blueward evolution, tend to undergo significantly more 
pulsations, even including radial strange-mode pulsations (with periods of 10-100 
or more days), as observed in $\alpha$ Cygni variables.

Additionally, the photospheric line profiles of BSGs typically contain
some significant extra broadening, so-called macroturbulence
\citepads{1958ApJ...127..658A, 1977ApJ...213..438C}, which is often on the
same order as rotational broadening \citepads{Ryans2002, Simon-Diaz2007, 
Markova2008}. Recent evidence links macroturbulence and line profile 
variability to stellar pulsational activity \citepads{Aerts2009, 
2010ApJ...720L.174S, 2010AN....331.1069S}, providing additional
proof for the presence of pulsations in BSGs.

BSGs lose mass via their line-driven winds, but the rates at which the 
material is lost are still highly
debated. Diverse spectral ranges trace distinct wind regions, and comparison
of mass-loss rates obtained from various methods typically results in values that
can differ by a factor 2--10. In addition, the observed complex wind structures
seen in the H$\alpha$ variability of BSGs
\citepads[e.g.,][]{1996A&A...305..887K, 2008A&A...487..211M},
for instance, imply that these
objects cannot have smooth spherically symmetric winds. 
Spectropolarimetric observations revealed no evidence of
magnetic fields \citepads{2014MNRAS.438.1114S}. Instead, the winds of BSGs are 
found to be clumped
\citepads{2010A&A...521L..55P}. Micro- and macro-clumping 
severely influence both mass-loss determination and line profile shapes 
of not only resonance lines,  but also H$\alpha$ 
\citepads[e.g.,][]{2010A&A...510A..11S, 2011A&A...528A..64S, 
2014A&A...568A..59S, 2012A&A...541A..37S, 2013A&A...559A.130S}.
The onset of clumping in the winds of massive stars could be due to 
line-driven instability \citepads[see, e.g.,][]{1998A&A...332..245F}. 
Perturbations leading to this instability might be initiated by pulsational 
activity, at least in the vicinity of the stellar photosphere.
Moreover, strange mode pulsations have been suggested to cause time-variable mass loss
in luminous, evolved massive stars \citepads[see][]{Glatzel1999,Aerts2009,2011IAUS..272..554P}. The first observational evidence of these modes was found 
in the massive BSG \object{HD\,50\,064} \citepads{Aerts2010}.

The ability of BSGs to maintain stable pulsations opens a completely new
perspective in studying these stars. Based on established methods from
asteroseismology, it will be possible to investigate not only the stellar
atmospheres from which the mass-loss and possible onset of wind clumping are
initiated, but also the deep interior of the stars, revealing important physical
properties, such as internal structure, rotation, and
mixing. Knowledge of these parameters is of vital importance for our understanding of
the post-main sequence evolution of massive stars, as these are key input
parameters to modern stellar evolution calculations.
This emphasizes the need for further investigations, both theoretically and
observationally, to improve our understanding of pulsational activity in BSGs
in different evolutionary stages.

To investigate, in particular, the possible interplay between pulsations and
mass-loss, we initiated an observational campaign to search for pulsational
activity and cyclic mass-loss variability in stellar winds for a sample of 
bright BSGs. Here we report on our results for the object 55\,Cyg.

\section{The star}
\label{sec:star}

55\,Cyg (HD\,198\,478, MWC\,353, HR\,7977, HIP\,102\,724, $\alpha$ = 
$20^{\mathrm{h}}\,48^{\mathrm{m}}\,56\,\fs 291$ and
$\delta$ = $\,+46^{\degr}\,06^{\arcmin}$ $50\,\farcs 88$; 2000) is a bright star 
{\rm located in the Cyg OB7 association \citepads{1978ApJS...38..309H}} that was
classified as B3\,Ia by \citetads{Morgan1950}. 
Its {\sc hipparcos} parallax of $1.40\pm 0.17$\,mas  \citepads{2007A&A...474..653V} 
places the star at a distance of $714\pm 86$\,pc.
The interstellar visual extinction toward 55\,Cyg was found to be 1.62\,mag 
\citepads{1977ApJ...213..737B,1978ApJS...38..309H}.

Studies of its photospheric lines revealed two important facts: different elements typically display
different radial velocities \citepads[e.g.,][]{Hutchings1970}, while the same lines show strong
variations with time. For instance, \citetads{1960PASP...72..363U} described radial velocity
variations of 25 km\,s$^{-1}$, with no evidence that the changes found were periodic. Later,
\citetads{1975A&A....45..343G} collected 34 spectra, distributed over 15 consecutive nights, 
and found that the radial velocity curves oscillate with a period of 4-5 days, in addition to
photospheric motions on timescales about three times longer.

Photometric observations are found to display microvariability, but clear periodicities are
 difficult to establish. While \citetads{1982A&AS...48..503R} reported an 18-day variability in their
V-band magnitudes, \citetads{2002MNRAS.331...45K} found a periodicity of 4.88 days using
{\sc hipparcos} data. On the other hand, no indication for clear periodicities were seen by
\citetads{1983PASP...95..491P}, \citetads{1989A&A...208..135V}, and \citetads{2009A&A...507.1141L}.

The H$\alpha$ line in  55\,Cyg is also highly variable. While \citetads{1960PASP...72..363U} and
\citetads{Rzaev2012} remarked that in their spectra the H$\alpha$ line maintained
a P\,Cygni profile, but with
highly variable strength, \citetads{1982ApJS...48..399E}  noted that during one of his observations,
H$\alpha$ showed additional faint emission wings, which were not visible anymore in the
spectrum he took two months later. Furthermore, \citetads{Maharramov2013} discovered
that H$\alpha$ can even completely disappear from the spectrum.

The stellar rotation velocity, projected on the line of sight ($v\sin i$), and the 
terminal wind velocity ($v_{\infty}$) are also highly variable or uncertain. For both, broad ranges are found
in the literature. Values for $v\sin\,i$ range from 0\,km\,s$^{-1}$ \citepads{1955ApJ...121..102S}
to a maximum of 61\,km\,s$^{-1}$ \citepads{1997MNRAS.284..265H}. The latter was obtained from
lines in the IUE spectra. Other determinations based on optical spectral lines and different
techniques include 42\,km\,s$^{-1}$ \citepads{1975MNRAS.173..419D}, $49\pm
1$\,km\,s$^{-1}$ \citepads{1992ApJ...387..673G}, $45\pm
20$\,km\,s$^{-1}$ \citepads{1999A&A...349..553M}, $35\pm 9$\,km\,s$^{-1}$ \citepads{Abt2002}, and
39\,km\,s$^{-1}$ \citepads{Markova2008}. The terminal wind velocity is much harder to determine.
Based on different lines in the IUE spectra, \citetads{1990ApJ...361..607P} obtained 470\,km\,s$^{-1}$
from the \ion{C}{iv} line, while \citetads{2010A&A...521L..55P} derived a higher value of
560\,km\,s$^{-1}$ from \ion{Si}{iv}. Both values are lower than
predictions from line-driven wind theory, however, which delivers a value of $\sim 690$\,km\,s$^{-1}$
\citepads[see, e.g.,][]{2001A&A...377..175K}.
Conversely, \citetads{Markova2008} noted that the observed value
of 470\,km\,s$^{-1}$ is too high to give satisfactory fits to their H$\alpha$ line, which
would agree much better with a terminal wind velocity of only 
$\sim 200$\,km\,s$^{-1}$.
These authors speculated that 55\,Cyg might have a time-variable wind velocity.

There is also scatter in the derived stellar parameters.
Although the determination of fundamental parameters could be biased by
the technique used, different values of $T_{\rm{eff}}$ are found using
models built on similar approaches. $T_{\rm eff}$ measurements based on the Balmer
discontinuity 
give values of 16\,450\,K \citepads[the color index method of][]{1992ApJ...387..673G}
and $15\,380\pm 860$\,K \citepads[the BCD spectrophotometric method of][]{2009A&A...501..297Z}, respectively. Using
nonlocal thermal equilibrium (non-LTE) plane-parallel hydrostatic stellar atmosphere and the Si
ionization balance, the derived $T_{\rm{eff}}$ values are 18\,000\,K
\citepads{1999A&A...349..553M}, 18\,500\,K \citepads{Monteverde2000}, and
17\,000\,K \citepads{Jurkic2011}, while non-LTE line- and wind-blanketed
model atmospheres provide values of 16\,500\,K
\citepads{2006A&A...446..279C} and 17\,500\,K
\citepads{2008A&A...481..777S, Markova2008}. In all cases, the
surface gravity determinations are around $\log\,g = 2.2\pm 0.1$
\citepads[cf.][]{Jurkic2011}.

As H$\alpha$ is a fundamental mass-loss diagnostic 
\citepads[e.g.,][]{1996A&A...305..171P}, it can be
expected that model fits to observations obtained at different periods (with
correspondingly varied shape and line strength) result in different mass-loss rates. And
the values clearly vary by a substantial amount. While \citetads{2008A&A...481..777S} found $\dot{M}
= 5\times 10^{-7}$\,M$_{\odot}$yr$^{-1}$, the value of \citetads{2006A&A...446..279C} is a factor of
two lower, even though both works used the same terminal wind velocity value of 470\,km\,s$^{-1}$.
In addition, \citetads{Markova2008} obtained a rather large uncertainty in their mass-loss values
$(1.175 - 4.07)\times 10^{-7}$\,M$_{\odot}$yr$^{-1}$, which is
due to the high uncertainty in the wind terminal
velocity (200\,km\,s$^{-1}$ vs 470\,km\,s$^{-1}$). All of these investigations used smooth 
(i.e., unclumped) winds.

Concerning the possible evolutionary stage of 55\,Cyg, 
\citetads{1969A&A.....1..494U} mentioned that the observed H$\gamma$ line 
profile was too shallow, which could indicate that 55\,Cyg is H-poor, and hence 
a post-red supergiant.
A similar conclusion was drawn by \citetads{1993A&AS...97..559L}, 
\citetads{1992ApJ...387..673G}, \citetads{2006A&A...446..279C}, and 
\citetads{2008A&A...481..777S} based on the CNO-processed material seen in its 
optical spectra, in particular, its strong N-enrichement. \citetads{Markova2008} 
additionally concluded that the star probably is enriched in He. These authors 
also found a rather low spectroscopic mass of $\sim 11$\,M$_{\odot}$ compared to 
the star's high luminosity, which places it on an evolutionary track of an 
initial mass of 25\,M$_{\odot}$. 
Additional support for its evolved nature comes from \citetads{1989A&A...208..135V}, who listed 55\,Cyg 
as an $\alpha$ Cygni variable (i.e., a luminous, photometrically variable supergiant of spectral type B and A). 
If this classification is correct, then 55\,Cyg 
should display numerous pulsation modes, including radial strange modes 
\citepads{Saio2013}.

\section{Observations}
\label{sec:obs}

We spectroscopically monitored 55\,Cyg
between 2009 August 15 and 2013 October 22. We obtained a total of 344 spectra,
distributed over 64 nights, with the Coud\'{e} spectrograph attached to the
Perek 2 m telescope at Ond\v{r}ejov Observatory \citepads{Slechta2002}. Until
the end of May 2013, the observations were taken with the 830.77 lines\,mm$^{-1}$
grating and a SITe $2030\times 800$ CCD. Beginning in June 2013, we used the 
newly installed PyLoN $2048\times 512$BX CCD. With both detectors, a
spectral resolution of $R\simeq 13\,000$ in the H$\alpha$ region was achieved,
and the wavelength coverages were from 6253\,\AA \ to  6764\,\AA \ for the old
CCD and from 6263\,\AA \ to  6744\,\AA \ with the new one. For wavelength calibration, a
comparison spectrum of a Th-Ar lamp was taken immediately after each exposure.
The stability of the wavelength scale was verified by measuring the wavelength
centroids of \ion{O}{i} sky lines. The velocity scale remains stable within
1\,km\,s$^{-1}$. Individual exposure times range from 400\,s to 2700\,s,  and, 
for all spectra, the achieved signal-to-noise (S/N) ratio was higher than
300. During the nights of 2013 July 19--22 and 2013 August 10--12, we achieved
S/N values of more than 500, allowing for high-quality time series, covering 
several hours of continuous observation.

We also obtained a series of echelle spectra using the Poznan Spectroscopic
Telescope at the Winer Observatory in Arizona, USA. Observations were carried out
from 2013 October 17 to 2013 November 7, and a total of 41 spectra were
acquired, covering the wavelength range 3900 to 8677\,\AA, \ with a spectral
resolution of $R = 40\,000$. Each spectrum has an exposure time of 1800\,s
and a S/N of $\sim 120-180$.

All data were reduced and the heliocentric velocity corrected using standard
IRAF\footnote{IRAF is distributed by the National Optical Astronomy
Observatories, which are operated by the Association of Universities for
Research in Astronomy, Inc., under cooperative agreement with the National
Science Foundation.} tasks. We also observed a rapidly rotating
star (HR\,7880, Regulus, $\zeta$\,Aql) once per night to perform the 
telluric correction.


\begin{figure}[t]
\begin{center}  
\includegraphics[width=\hsize,angle=0]{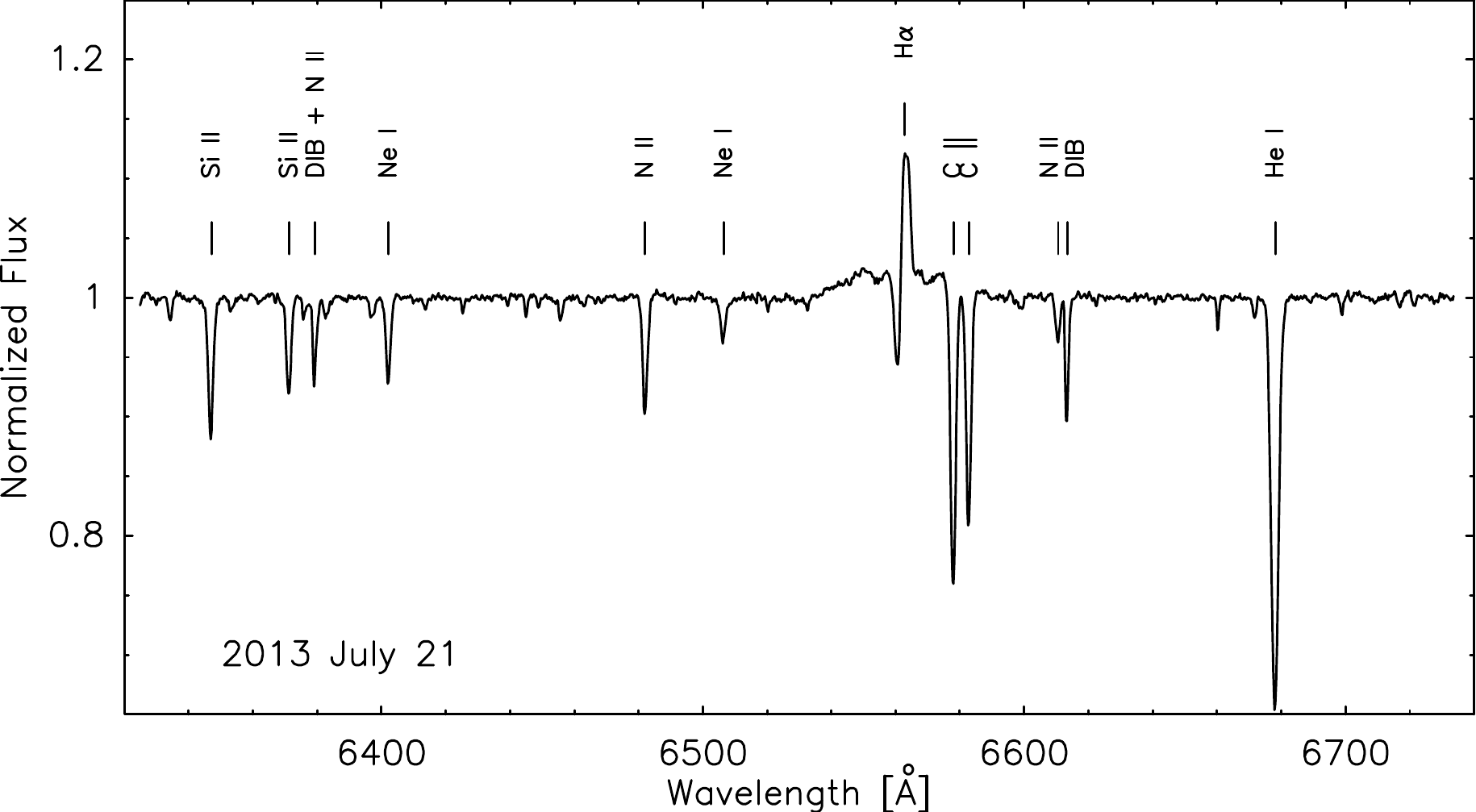}
\caption{Spectrum of 55\,Cyg taken in the H$\alpha$ region with the Perek
2m telescope at Ond\v{r}ejov Observatory.}
\label{fig:spectrum}
\end{center}
\end{figure}


\begin{figure}[t]
   \centering  
   \includegraphics[width=\hsize,angle=0]{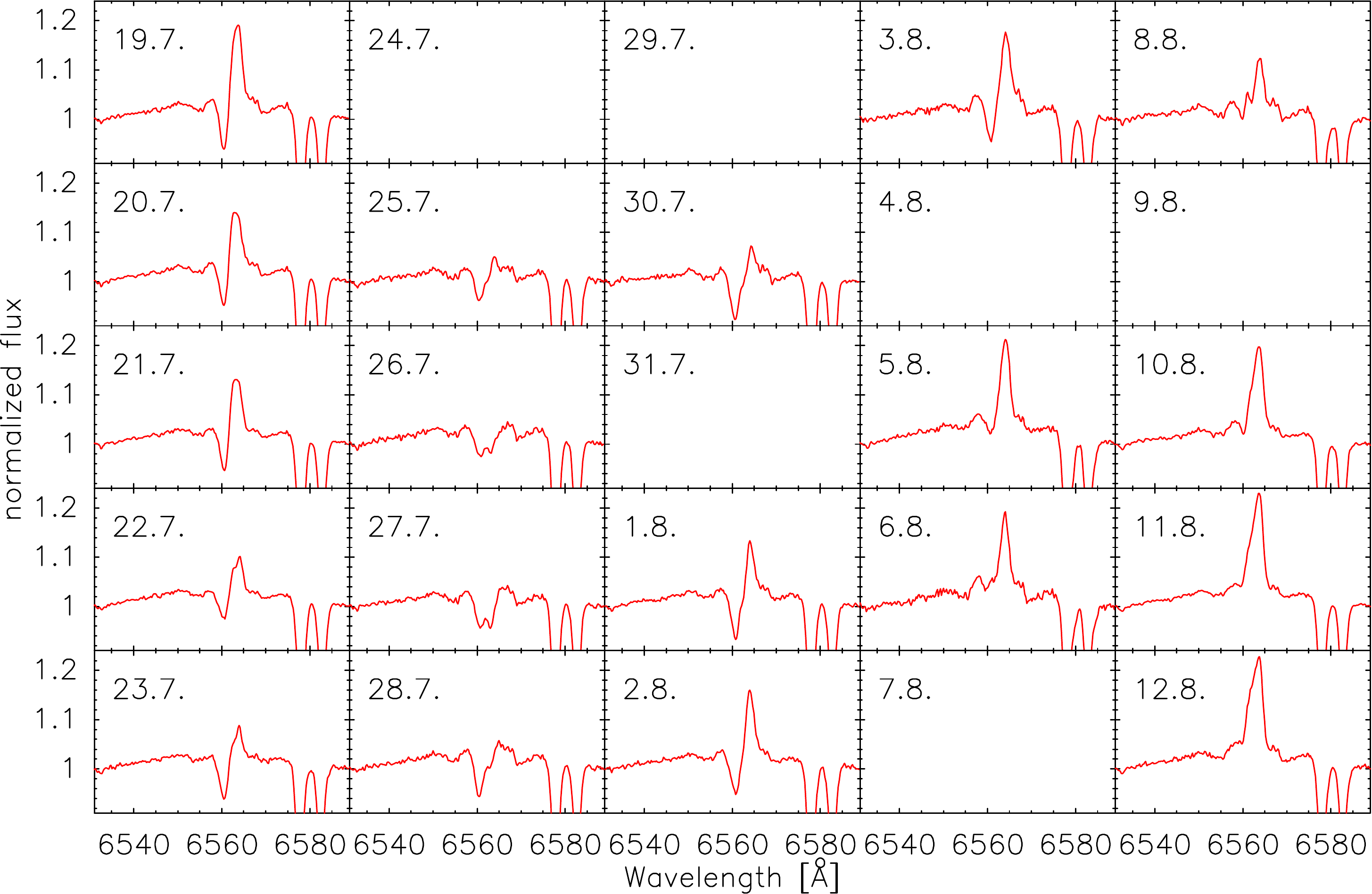}
      \caption{Night-to-night variation in the H$\alpha$ line in the
        observing period 2013 July 19 to August 12. This period shows all
of the typical profile shapes: P\,Cygni, pure emission, almost complete absence,
and double and multiple peaked.}
         \label{fig:halpha}
\end{figure}

A typical Ond\v{r}ejov spectrum is shown in Fig.\,\ref{fig:spectrum}. In 
general, the overall appearance of the full optical spectrum is very similar 
to the ones presented by \citetads{1992A&AS...94..569L} and 
\citetads{2007AstBu..62..257C}. Interesting is the shallow, but broad emission 
component centered on the H$\alpha$ line, which is noticeable in each of our 
spectra,
while it was reported only once in the literature 
\citepads{1982ApJS...48..399E}. The extent of the 
emission wings measured from our spectra is $\pm 1200$\,km\,s$^{-1}$. Such 
broad emission components typically originate from incoherent electron 
scattering \citepads[see, e.g.,][]{1997ApJ...482..757M,1999A&A...350..970K}.

The H$\alpha$ line displays time variations in both intensity and shape, 
showing various types of profiles: P\,Cygni, pure emission, almost complete 
absence, and double or multiple peaked. As an example, see Fig.\,\ref{fig:halpha}.
Similar variability of the H$\alpha$ line  has been observed in other BSGs 
that belong to the group of $\alpha$ Cygni variables \citepads[e.g.,][]{1996A&A...305..887K}.

\section{Results}
\label{sec:results}

\subsection{Line profile modeling}

To quantify and analyze the spectral line variation of 55\,Cyg, we 
simultaneously fit observed line profiles of several representative transitions 
of H, He and Si atoms 
with synthetic line models computed with the code Fast Analysis of 
STellar atmospheres with WINDs (FASTWIND). This is a non-LTE, spherically symmetric 
model atmosphere code that enables computating continuum and line radiation
fluxes and provides a set of parameters that describe the photospheric and 
wind structure \citepads{1997A&A...323..488S,2005A&A...435..669P}.
The code includes line blocking and blanketing effects and consistently calculates 
the temperature structure \citepads[see details in][]{Kubat1999}.
The wind velocity distribution is computed using the so-called $\beta$-law.

For the fundamental stellar parameters ($T_{\rm eff}, \log\,g$)
we mainly used the Arizona spectra. They cover the H$\gamma$ and H$\delta$
lines (known as excellent surface gravity sensors) and the
\ion{He}{i}\,$\lambda$\,4471 triplet line \citepads[considered a good
gravity indicator, with some sensitivity to $T_{\rm eff}$ as well,][]{Lefever2007}. 
In addition, we modeled the singlet transitions of 
\ion{He}{i}\,$\lambda$\,6678\,\AA \ and \ion{He}{i}\,$\lambda$\,4713\,\AA, and
the lines of \ion{Si}{ii}\,$\lambda\lambda$\,4128,\,4130\,\AA, and 
\ion{Si}{iii}\,$\lambda$\,4552\,\AA. These Si lines are also excellent 
temperature indicators \citepads{Markova2008}.
Unfortunately, the Ond\v{r}ejov spectra cover a more restricted wavelength interval, providing only the \ion{He}{i}\,$\lambda$\,6678\,\AA \ line. Although the 
Ond\v{r}ejov spectra have a lower resolution, they have significantly higher
S/N ratios than the Arizona spectra.

Another input model parameter is the stellar radius. This value should be 
derived in an independent way, such as the de-reddened absolute magnitude or 
the spectral energy distribution. Different authors used these methods to 
derived the radius of \object{55 Cyg} and obtained values ranging from 
$38\,R_\sun$ \citepads{Markova2008} to $83\,R_\sun$ 
\citepads{1992ApJ...387..673G}. Because of the large discrepancy in the literature 
values, we employed other methods. Using the BCD parameters, based on a 
reddening-independent method \citepads{2009A&A...501..297Z}, we obtain a 
bolometric magnitude of $-8.5\pm 0.3$\,mag, which leads to $\log\,L/L_\sun = 
5.3\pm 0.1$ and $R = 61\pm 8\,R_\sun$. This bolometric magnitude agrees with 
the determination reported by \citetads{1977ApJ...213..737B}, who estimated 
$M_{\rm V} = -6.8$\,mag from the 10\,$\mu$m excess emission. With the
bolometric correction of 1.7 \citepads{Flower1996}, we derive $M_{\rm{bol}} = 
-8.6$\,mag, in agreement with the BCD method.
The radius obtained from these two methods also agrees with the value of
61\,R$_\odot$, derived by \citetads{2001A&A...367..521P} based on
measurements of the stellar angular diameter from infrared photometry by
\citetads{1977MNRAS.180..177B}.

The wind parameters were derived by modeling the H$\alpha$ line.
To compute the synthetic line profiles for 55\,Cyg, we started 
from a set of initial values for both photospheric and  wind 
parameters, taken from the literature (see Sect. \ref{sec:star}). However, these 
values did not deliver good fits, so that we extended the range 
of parameters, looking for the best fit in a ``by-eye'' procedure. For 
this, we treated all model parameters ($T_{\rm {eff}}$, $\log\,g$, $\beta$, 
$\dot{M}$, $v_{\infty}$, $v_{\rm{macro}}$, and $v_{\rm{rot}}$) as 
free values and computed a large grid of synthetic line profiles.
The parameters $v_{\rm{micro}}$ and $R_{*}$ were initially fixed at 
10\,km\,s$^{-1}$ and 61\,R$_\odot$ , respectively, and then were
slightly varied to 
obtain improved fits. We selected the best fit to the observations and 
verified the consistency of some of the individual parameters, which are not 
independent variables. For example, the stellar radius and the surface 
gravity should obey the relation $g\,R_{*}^{2} \approx $ constant. 
To fit the terminal velocity, we
preferentially considered the values that reproduce the blueshifted 
absorption component of the H$\alpha$ P\,Cygni profile. Otherwise,
for profiles with pure emission, we derived the terminal velocity based 
mainly on the H$\alpha$ line width.

Using the FASTWIND code, we explored the sensitivity of the
model parameters that characterize both the photosphere and the wind. 
However, we should keep in mind that there are two sources of errors. 
One results from the modeling, in which the involved uncertainties result from 
the statisitcal standard deviation and are obtained using the best-fit model.
The second refers to the inaccuracy of the input parameters, that is, the radius,
resulting from the uncertainties in the stellar distance and hence luminosity.
But when the distance is fixed, any deviation in the parameters larger 
than the statistical error indicates possible variability in the physical 
parameters. 

To determine the statistical uncertainty of the parameters, we used the best-fit model and tested different values of each
parameter, one at a time, keeping all other parameters fixed. We have the
advantage of the many high-resolution and high S/N
ratio spectra obtained in Arizona, taken over a sequence of consecutive
days. This enabled us to distinguish errors on the mean values of the
studied parameters from real changes of the physical conditions.

From fits to the \ion{Si}{ii} and \ion{Si}{iii} lines we observe 
uncertainties of 300 K - 500 K in  $T_{\rm eff}$.
We also find that the H$\alpha$ emission is sensitive to changes in 
$T_{\rm eff}$ of $\sim 300$\,K. From the H$\gamma$ and H$\delta$ wings,
the statistical error in $\log\,g$ is 0.05 dex, but due to large 
uncertainties in other parameters (e.g., the radius), the error could be 
slightly larger, but still below 0.1\,dex. This latter is what we use in the tables as 
upper limit.
The uncertainties in the 
rotational, micro- and macroturbulent velocities depend on the modeled line 
transition. Error bars in $v_{\rm micro}$ are 2\,km\,s$^{-1}$ for the He and 
Si lines, and 5\,km\,s$^{-1}$ for the H lines,
except for H$\alpha$, for which the error is about 10\,km\,s$^{-1}$. 
We find errors of 5\,km\,s$^{-1}$ in $v_{\rm rot}$ and
$v_{\rm macro}$ for H$\gamma$, H$\delta$, He, and Si lines, and of 
5\,km\,s$^{-1}$ and 10\,km\,s$^{-1}$ for H$\alpha$. For H$\beta$, our
error estimates for $v_{\rm rot}$ and $v_{\rm macro}$ are $\sim 20$\,km\,s$^{-1}$.
To derive the wind parameters, we considered the shape and intensity of
the absorption and emission components of the H$\alpha$ line. 
We find noticeable line variations with changes in $\dot{M}$ of about 
$0.1 \times 10^{-7}$M$_\odot$\,yr$^{-1}$ and 10\% in $v_{\infty}$ (although 
the error
in $v_{\infty}$ can reach up to 30\% if H$\alpha$ displays a pure absorption
line). 
These adopted uncertainties are overestimations because they are larger than or equal to the standard statistical 
deviation computed with the model parameters of the Arizona data, which 
correspond to a term of 21 days showing small or moderate spectral variations.
Unfortunately, even if we can check the sensitivity of  $\dot{M}$  by varying the parameters 
of the model, we cannot provide an accurate value of this quantity because
this depends on the uncertainty of the stellar distance. To evaluate the 
error in relation to the accurate value, we performed an error propagation using 
the optical depth parameter $Q= \dot{M} / (v_{\infty}R_{*})^{1.5}$, which 
results from the model calculations. To compute $Q,$ we selected the 
observations of Arizona (listed in Table\,\ref{tab2:results}, $\log Q = -12.975\pm 0.066$)
because they were taken during many 
consecutive days and the line profiles did not show huge variations. Assuming 
the scatter in the bolometric
magnitude $\Delta M_{\rm bol} \simeq \pm 0.3$, that is, $\Delta \log L/L_{\odot} 
\simeq 0.12$, and the error in $T_{\rm eff} \simeq 500$\,K, we obtain $\Delta 
\log R_{*}/R_{\odot} \simeq 0.04$. With our uncertainty of 10\% in $v_{\infty}$
and the statistical error $\Delta \log Q \simeq \pm 0.07$, the uncertainty in 
$\Delta\dot{M}/\dot{M} \simeq 0.3$.

The best-fit values (and corresponding statistical deviations) for the 
photospheric and wind parameters derived for each observation are given in 
Tables \ref{tab2:results} (Arizona data) and  \ref{tab1:results} (Ond\v{r}ejov data).
We adopt solar abundances for all elements, but discuss the validity, in 
particular with respect to the He abundance, in Sect.\,\ref{chemical}.

\begin{figure}[t]
   \centering  
   \includegraphics[width=\hsize,angle=0]{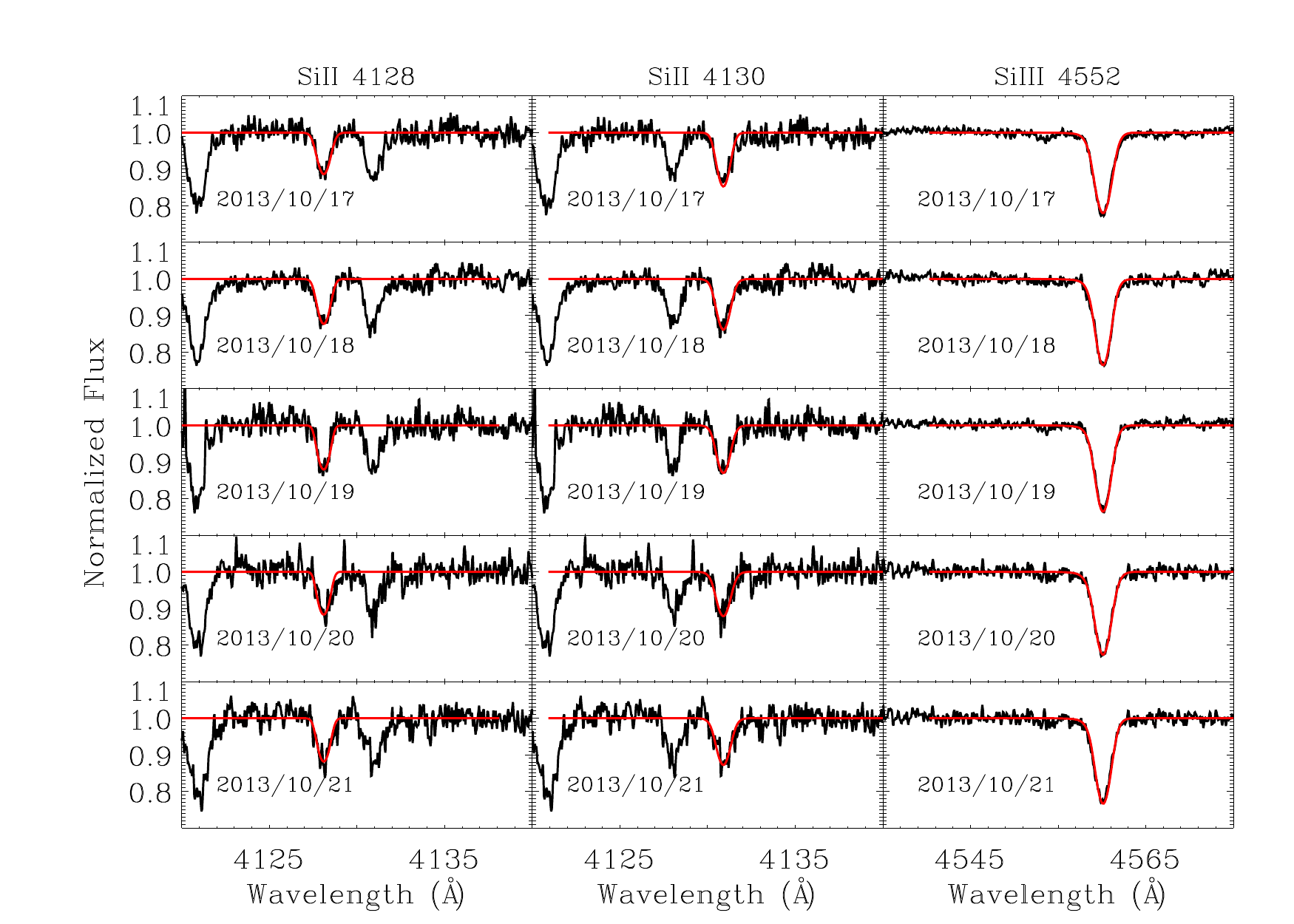}
   \includegraphics[width=\hsize,angle=0]{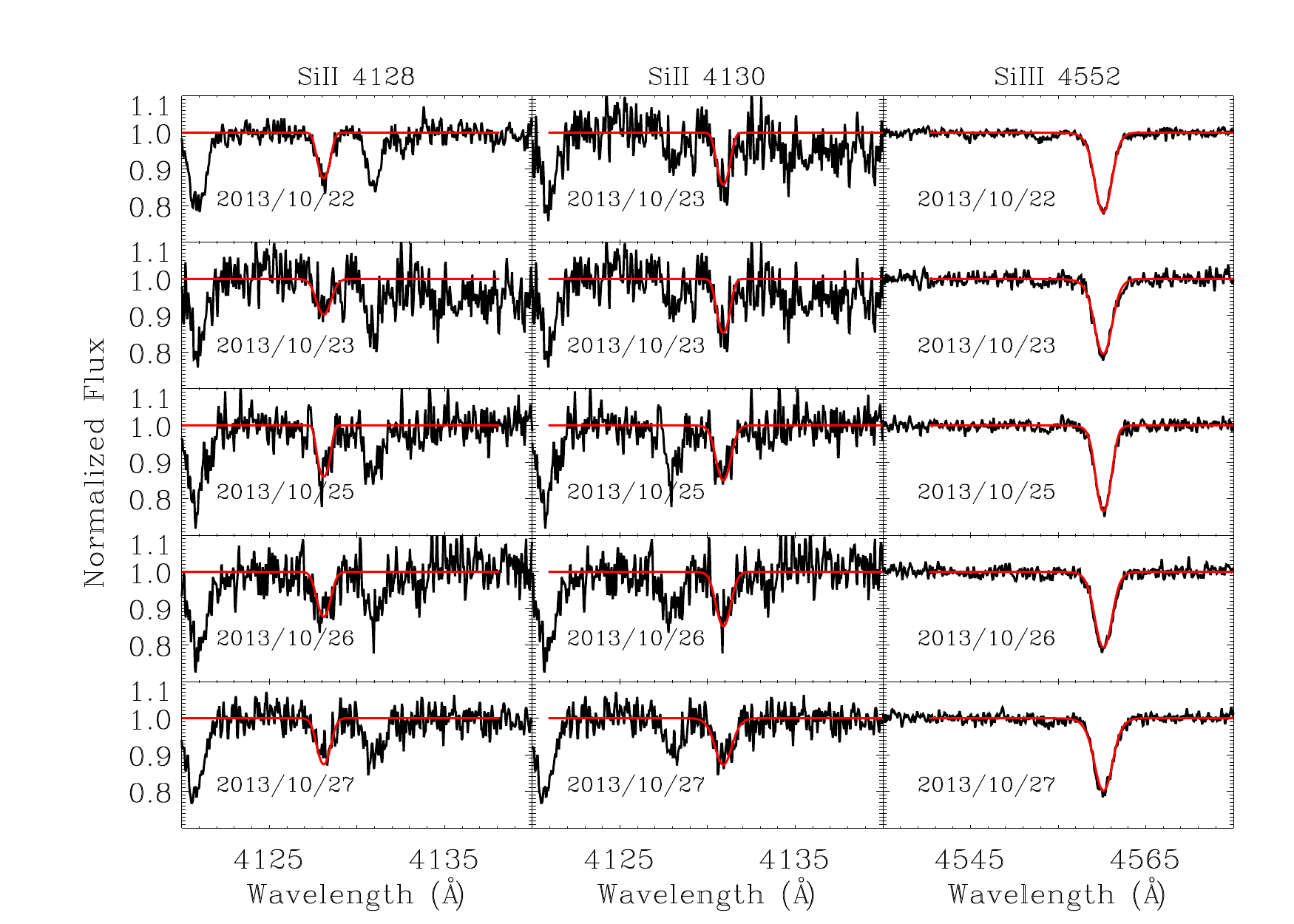}
   \includegraphics[width=\hsize,angle=0]{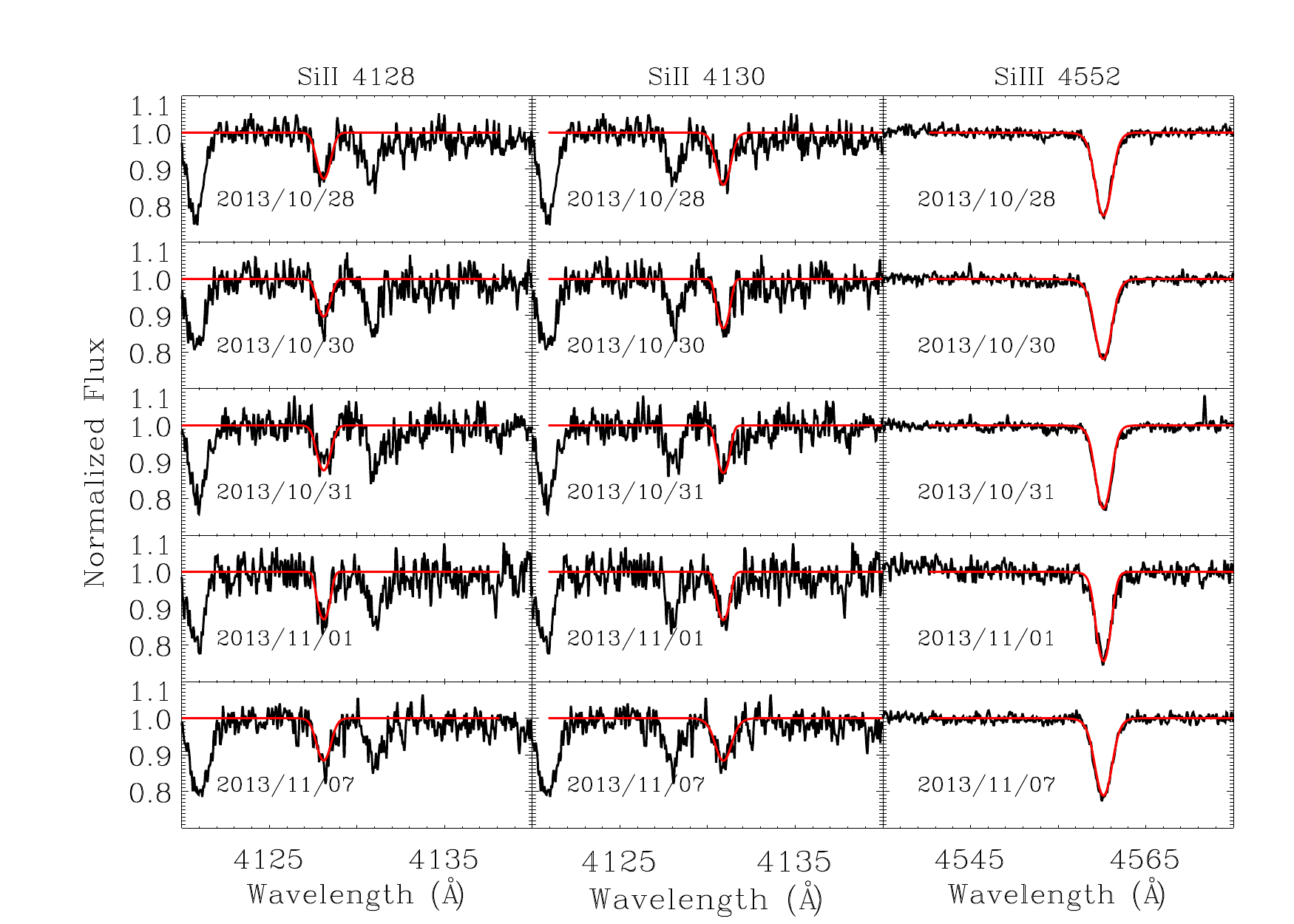}
      \caption{Fits to the lines of \ion{Si}{ii}\,$\lambda\lambda$\,4128,\,4130 \AA, and \ion{Si}{iii}\,$\lambda$\,4552\,\AA\, in the Arizona spectra.
Fits to the different line transitions are achieved using individual values of
$\varv_{\rm{mic}}$ and $\varv_{\rm{mac}}$ (for details see Sect.\,\ref{vel}).}
         \label{fig:Si}
\end{figure}

Figure \ref{fig:continued:one} shows a sequence of model fits to 
simultaneous observations of H$\alpha$, H$\beta$, H$\gamma$, H$\delta$, 
\ion{He}{i}\,$\lambda$\,6678\,\AA, \ion{He}{i}\,$\lambda$\,4471\,\AA, \ and 
\ion{He}{i}\,$\lambda$\,4713\,\AA\, line spectra taken in Arizona during 15 
successive nights in 2013  (see also Table \ref{tab2:results}). The
fits to the lines of \ion{Si}{ii}\,$\lambda\lambda$\,4128,\,4130 \AA, and \ion{Si}{iii}\,$\lambda$\,4552\,\AA\, are shown in Fig.\,\ref{fig:Si}.
We obtained 
very good fits for the whole sample of lines, with the exception of the blue 
wing of \ion{He}{i}\,$\lambda$\,4471\,\AA, where we observe a tiny 
discrepancy between the model and the spectrum. This discrepancy 
corresponds to the presence of a synthetic weak \ion{He}{i} forbidden component 
at $\lambda$\,4470\,\AA, which is not observed in any of our spectra. 
H$\alpha$ shows variations in strength in both the absorption and emission 
components of the P Cygni profile. 
Interestingly, the strength and width of the \ion{Si}{ii} lines shows 
night-to-night variations, while the \ion{Si}{iii} lines remain almost 
constant. The obtained $\log$ 
(EW\,({\ion{Si}{ii}\,$\lambda$ 4128})/EW\,({\ion{Si}{iii}\,$\lambda$ 4553})) 
ranges between -0.3 and -0.5. This variation can be attributed to a small change 
in the ionization rate of \ion{Si}{ii}.

\begin{figure*}[t!]
\setlength\fboxsep{0pt}
\setlength\fboxrule{0.5pt}
\begin{center}  
\includegraphics[width=\linewidth]{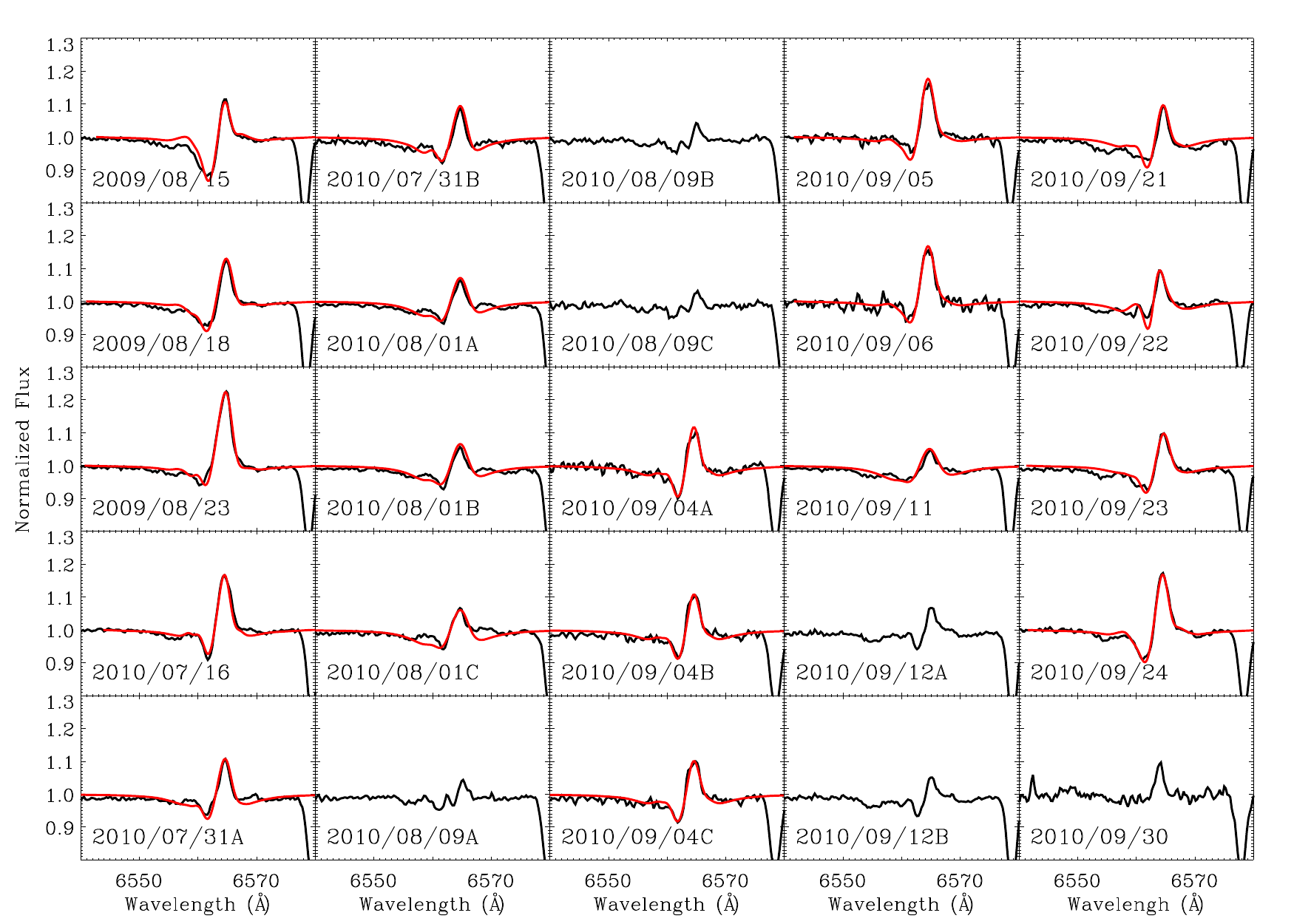}
\caption{H$\alpha$ line profiles taken in 2009 and 2010 at Ond\v{r}ejov Observatory together
with their corresponding model fits (for details see Table \ref{tab1:results}). Lines with highly complex profiles or roughly zero EW have not been fitted.
\label{halpha2009}}
\end{center}
\end{figure*}

\begin{figure}
\setlength\fboxsep{0pt}
\setlength\fboxrule{0.5pt}
\begin{center}  
\includegraphics[width=\linewidth]{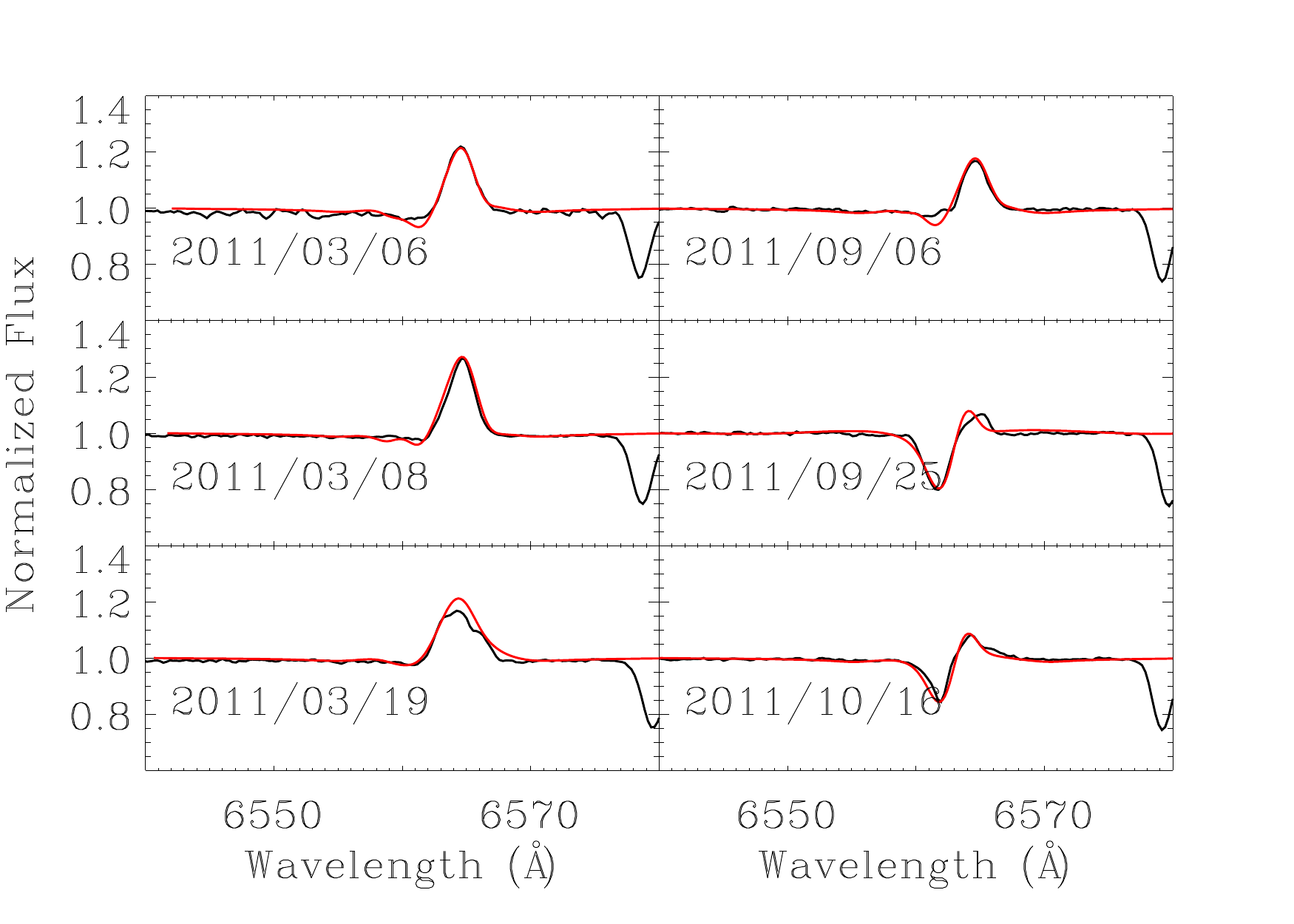}
\caption{Same as Fig.\,\ref{halpha2009}, but for data taken in 2011.}
\label{halpha2011}
\end{center}
\end{figure}

\begin{figure*}[t!]
\setlength\fboxsep{0pt}
\setlength\fboxrule{0.5pt}
\begin{center}  
\includegraphics[width=.80\hsize]{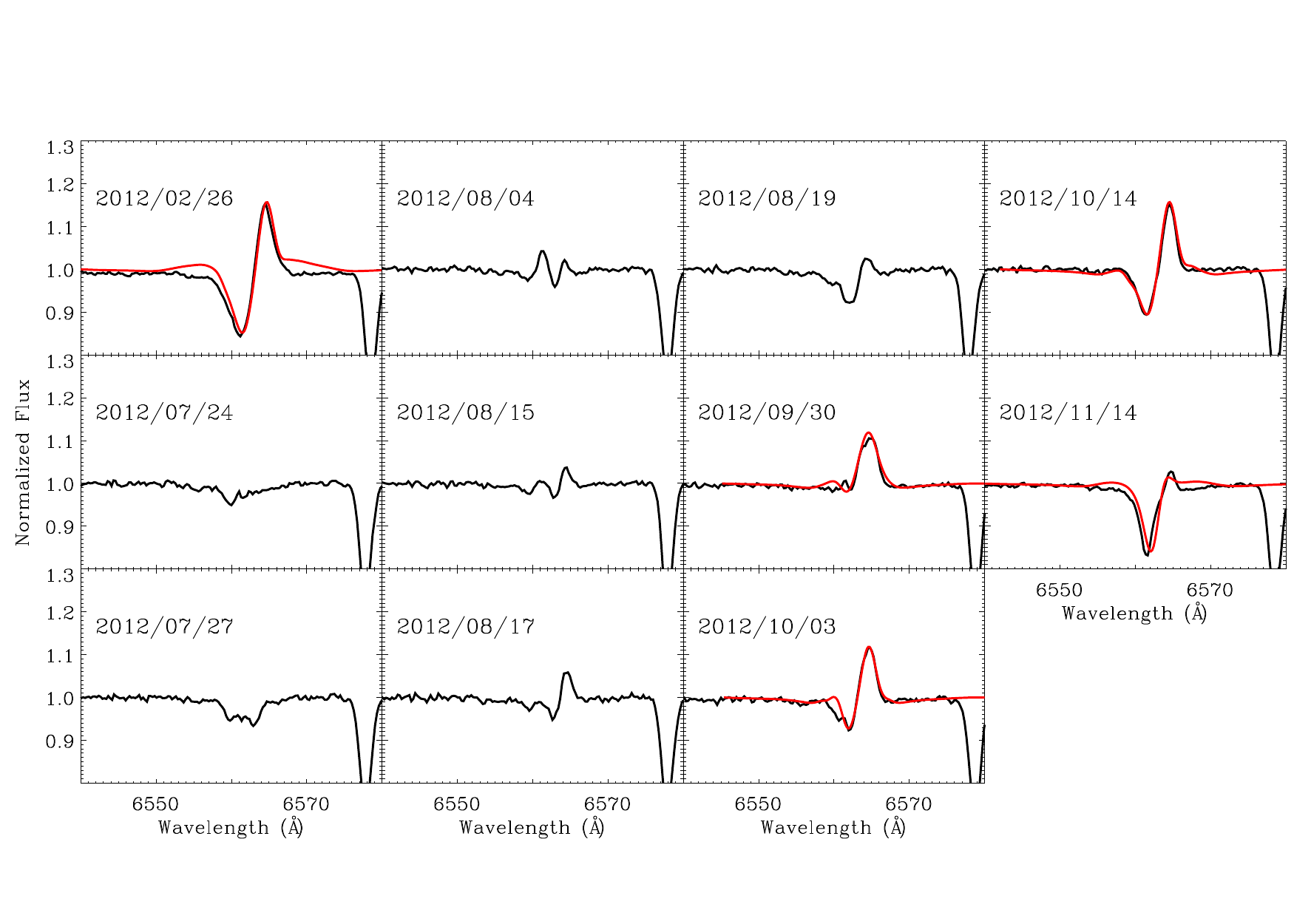}
\includegraphics[width=.80\hsize]{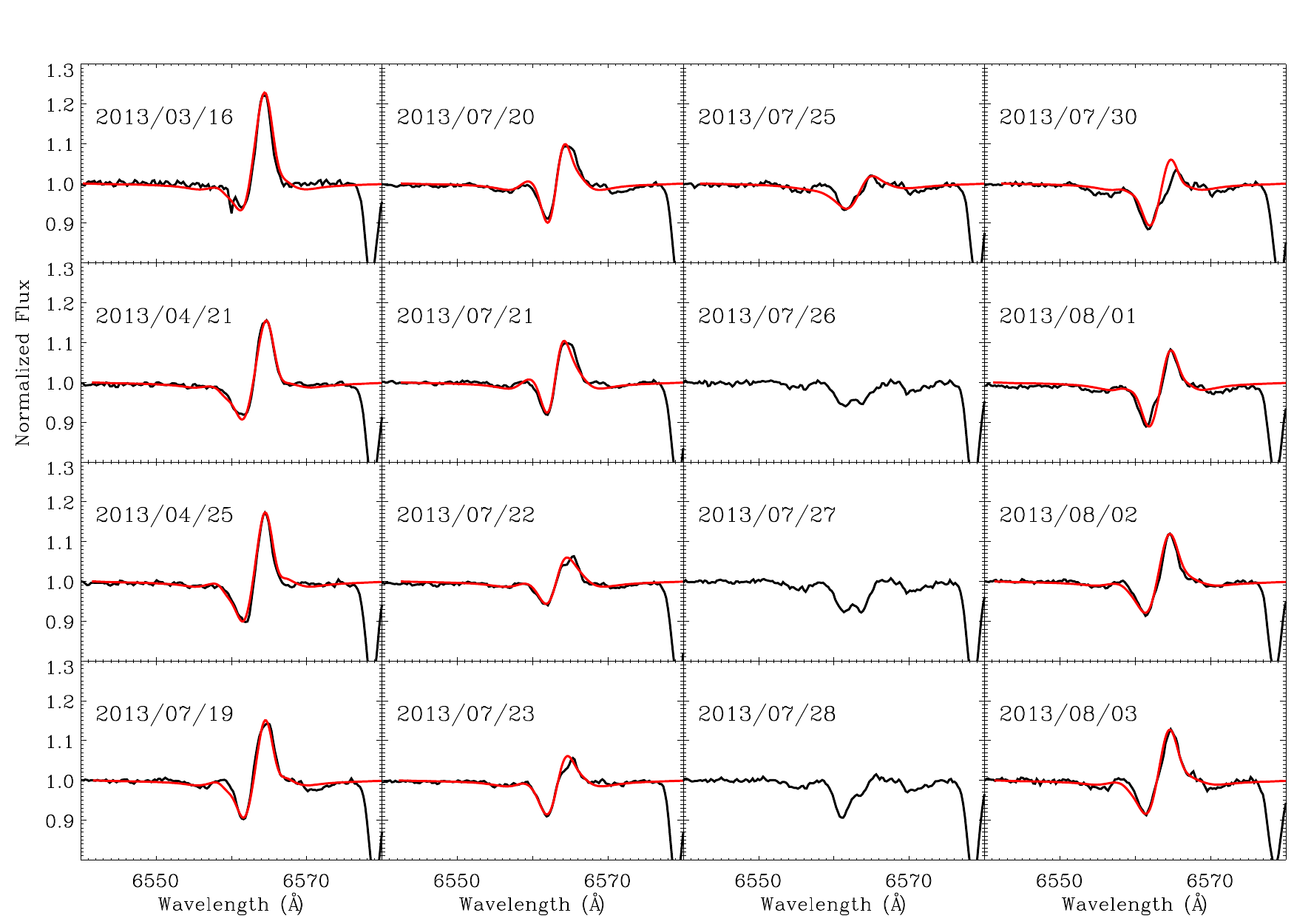}
\caption{Same as Fig.\,\ref{halpha2009}, but for data taken in 2012 (top) and 2013 (bottom).}
\label{halpha2012}
\end{center}
\end{figure*}

Fits to the time-scattered observations taken in 
Ond\v{r}ejov are displayed in Figs.\,\ref{halpha2009}--\ref{halpha2012}. 
As these spectra contain only two lines that can be modeled, that is, 
H$\alpha$ and \ion{He}{i}\,$\lambda$\,6678\,\AA, the scatter in the obtained 
values is slightly larger. In general, the models match the H$\alpha$ line 
profile  very well, with the exception of those observations displaying complex 
features in absorption  and/or double or multiple peaked emissions. We note an 
interesting event of enhanced mass-loss, observed on 2011 September 25, when
the H$\alpha$ line shows an unusual P-Cygni profile with a deep 
absorption component and a small emission feature  (see Fig.\,\ref{halpha2011}).


\begin{figure*}[tbp]
  \begin{center}
\includegraphics[width=.9\linewidth]{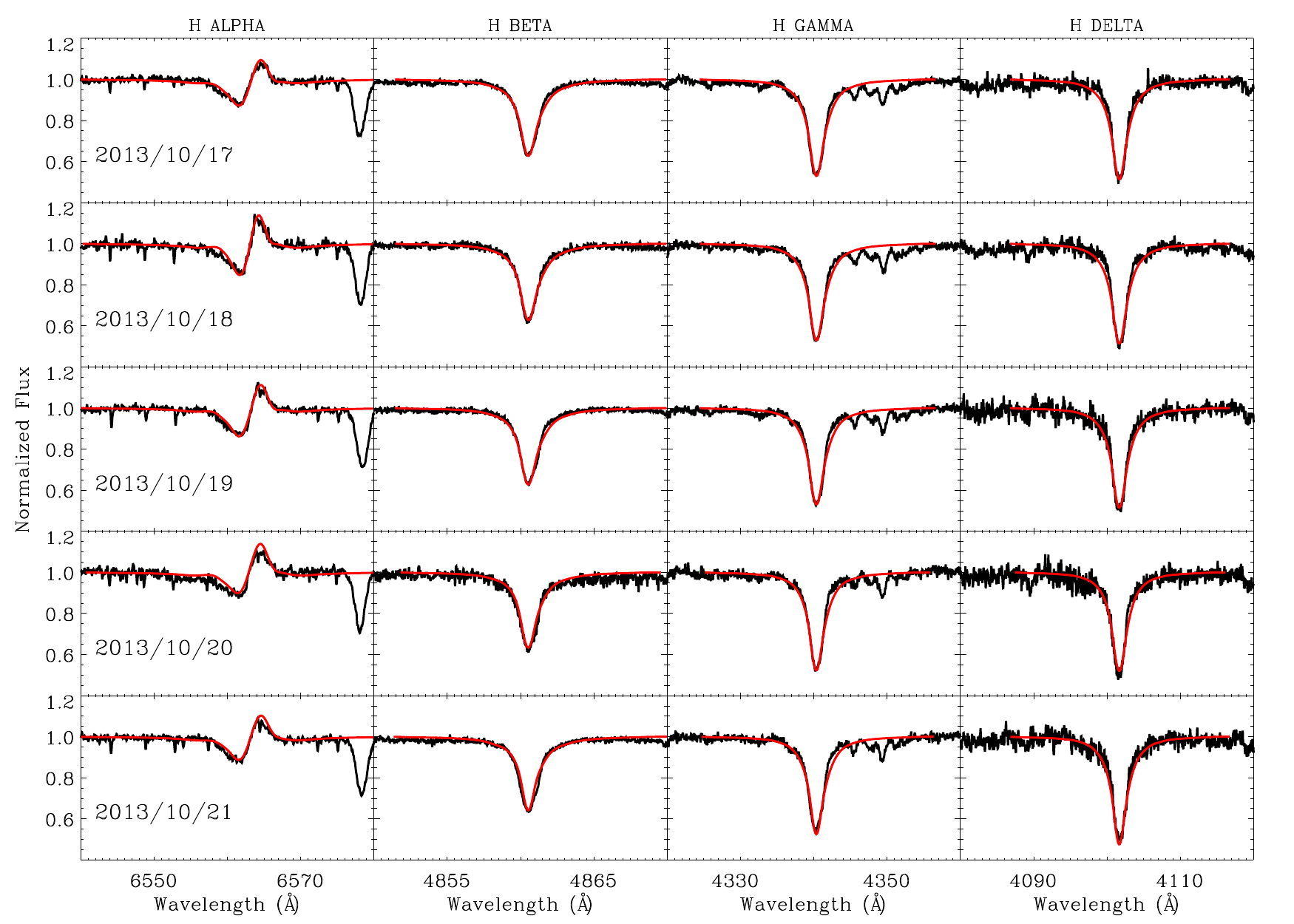}
\hspace*{0.5cm}
 \includegraphics[width=.82\linewidth]{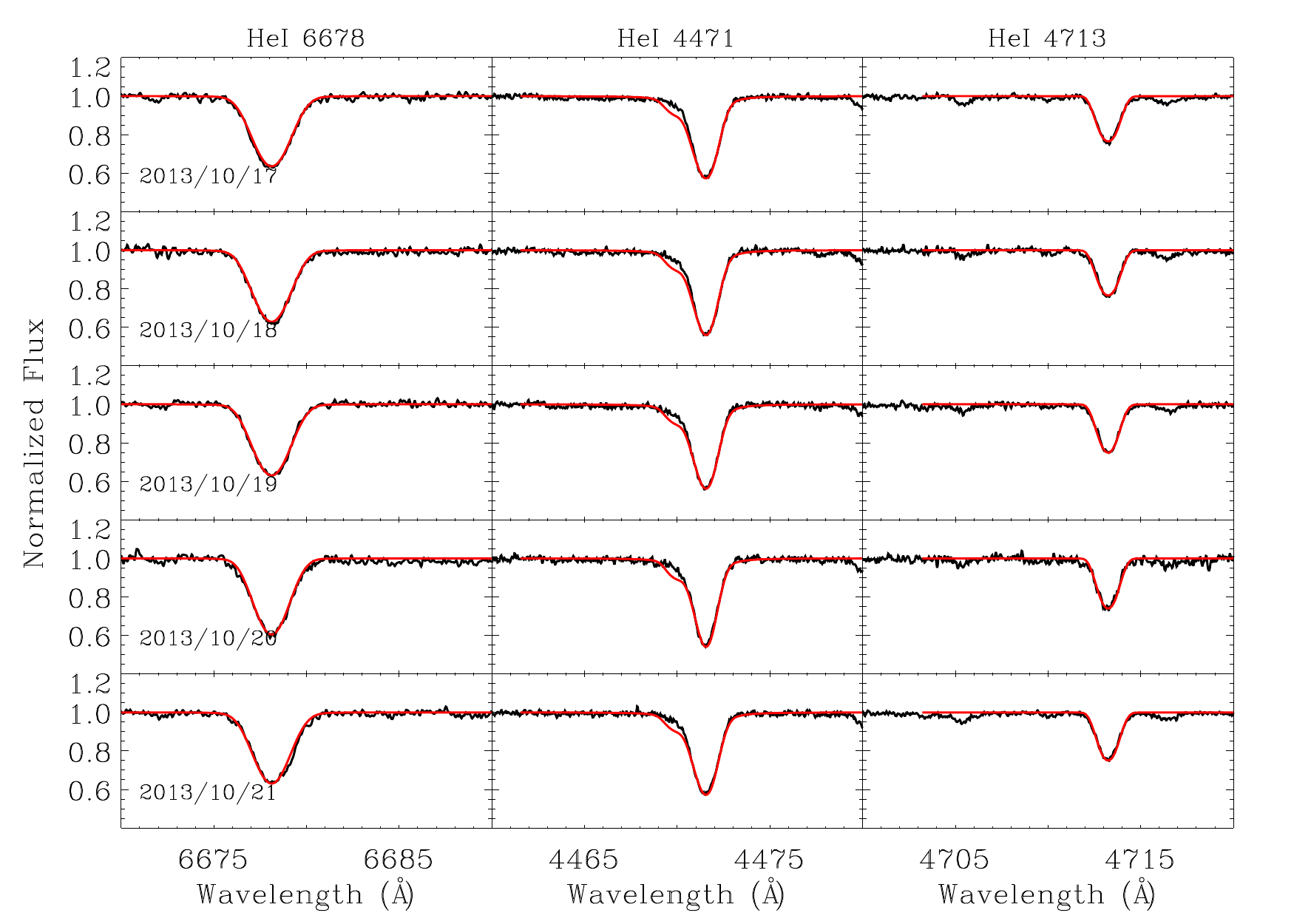}
\hspace*{0.5cm}
\caption{Sequence of fits to daily line variations of H and He lines in 2013.
Fits to the different line transitions are achieved using individual values of
$\varv_{\rm{mic}}$ and $\varv_{\rm{mac}}$ (for details see Sect.\,\ref{vel}).\label{fig:continued:one}}
\end{center}
\end{figure*}

\begin{figure*}[tbp]
 \ContinuedFloat
 \begin{center}
        \label{subfig:continued:subtwo}
 \includegraphics[width=.90\linewidth]{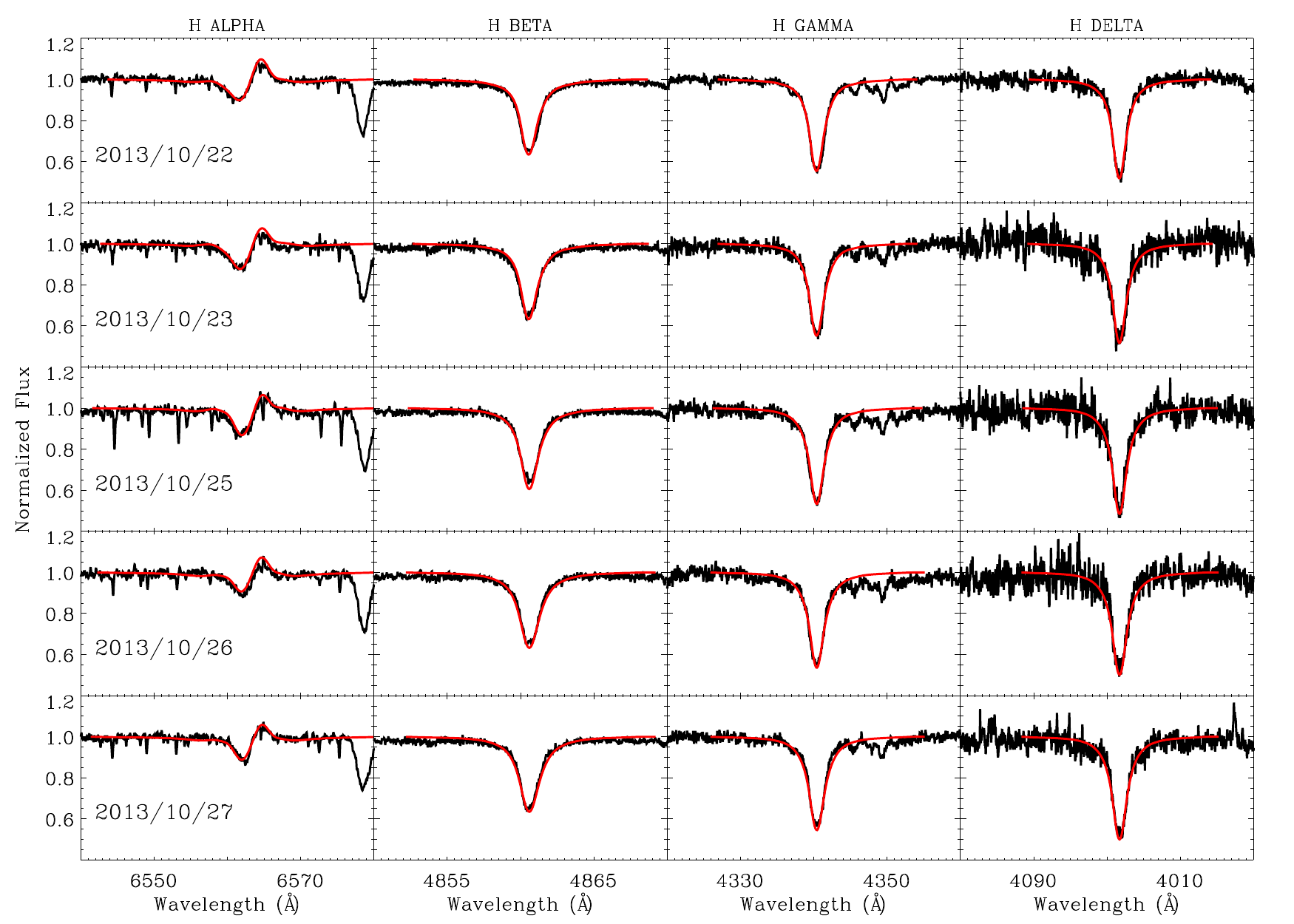}
\hspace*{0.5cm}
 \includegraphics[width=.82\linewidth]{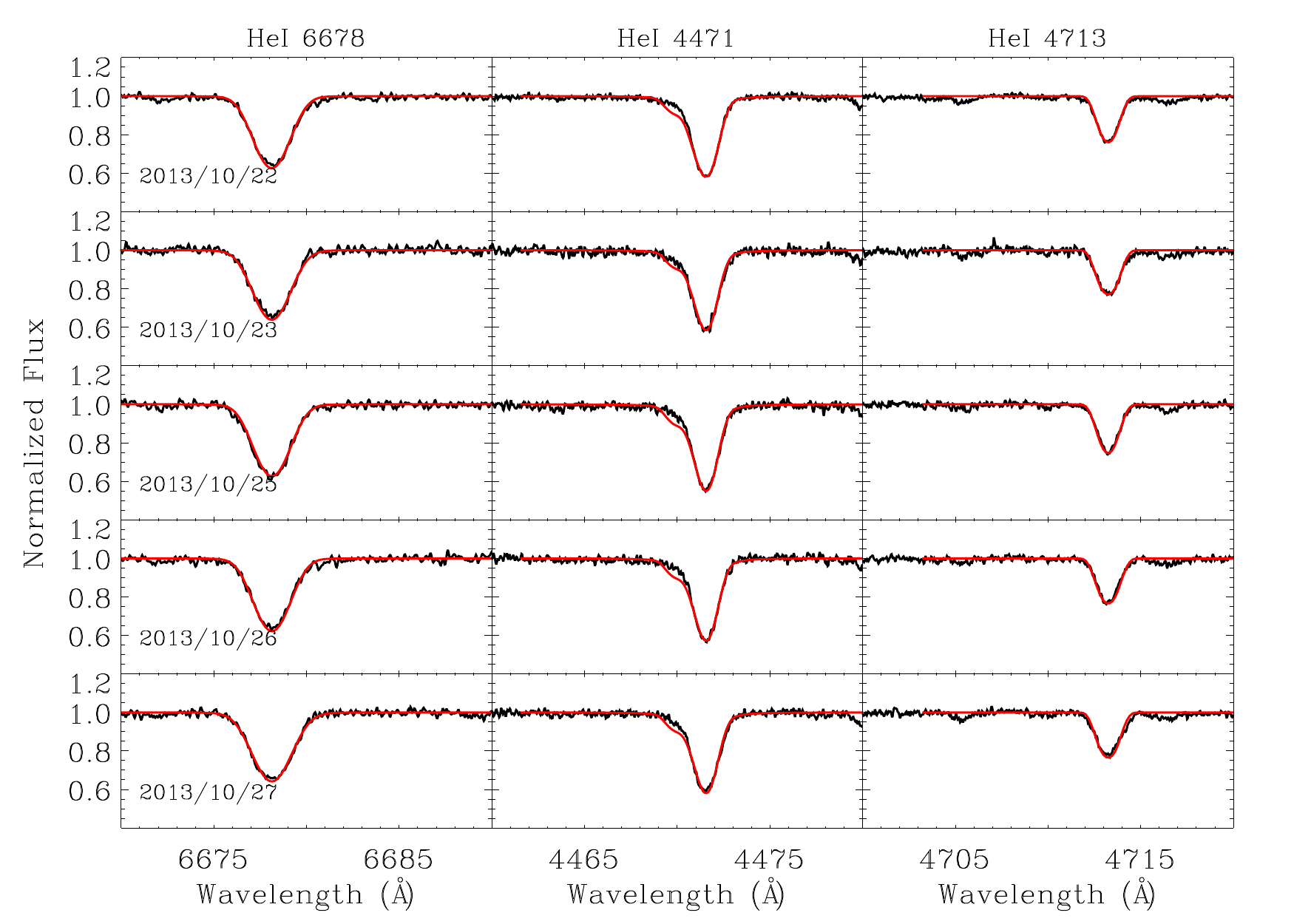} \caption{Continued.}

\hspace*{0.5cm}
\label{fig:continued:two}
 \end{center}
\end{figure*}


\begin{figure*}[tbp]
 \ContinuedFloat
 \begin{center}
        \label{subfig:continued:subthree}
\includegraphics[width=.90\linewidth]{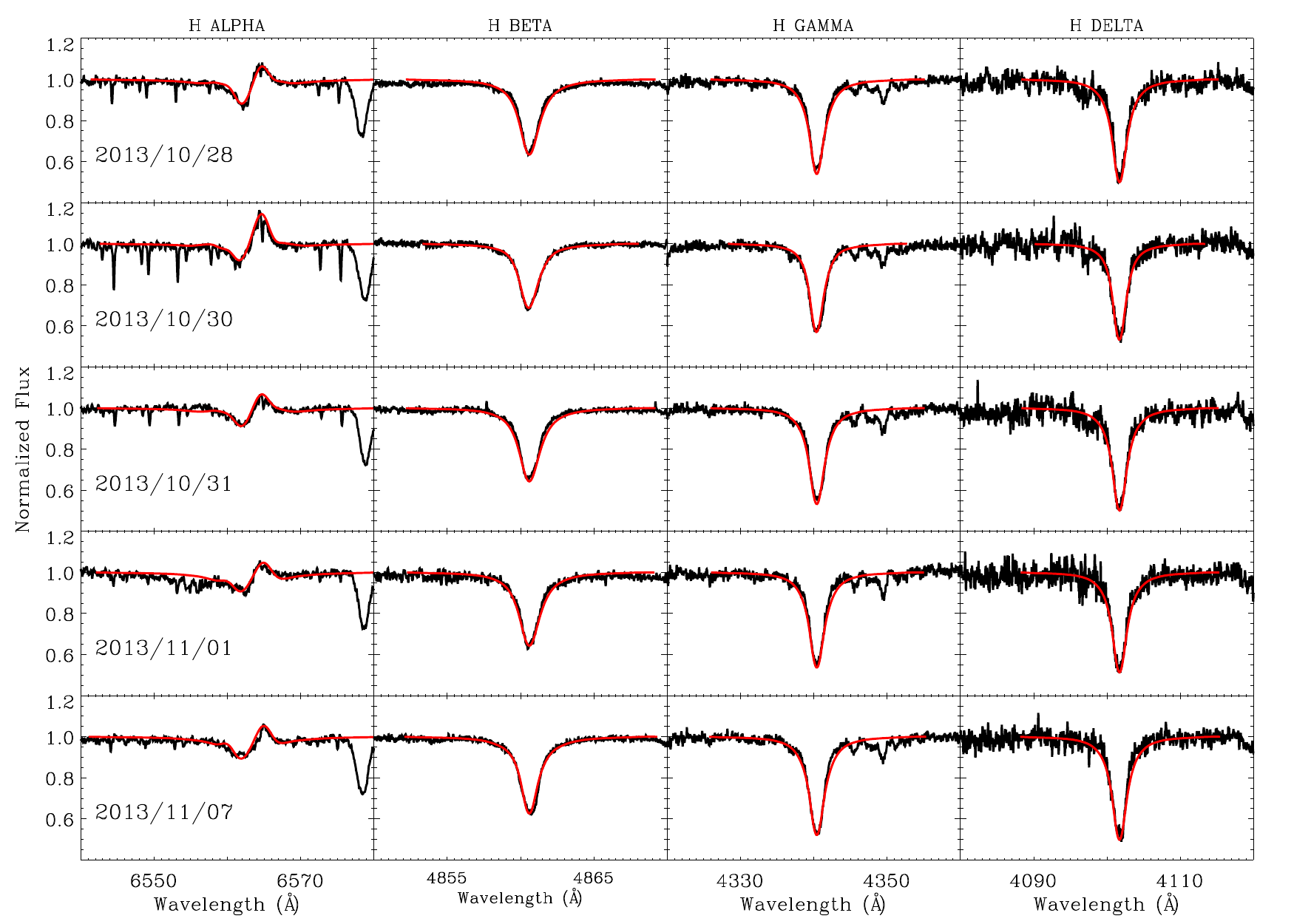}
\hspace*{0.5cm}

 \includegraphics[width=.82\linewidth]{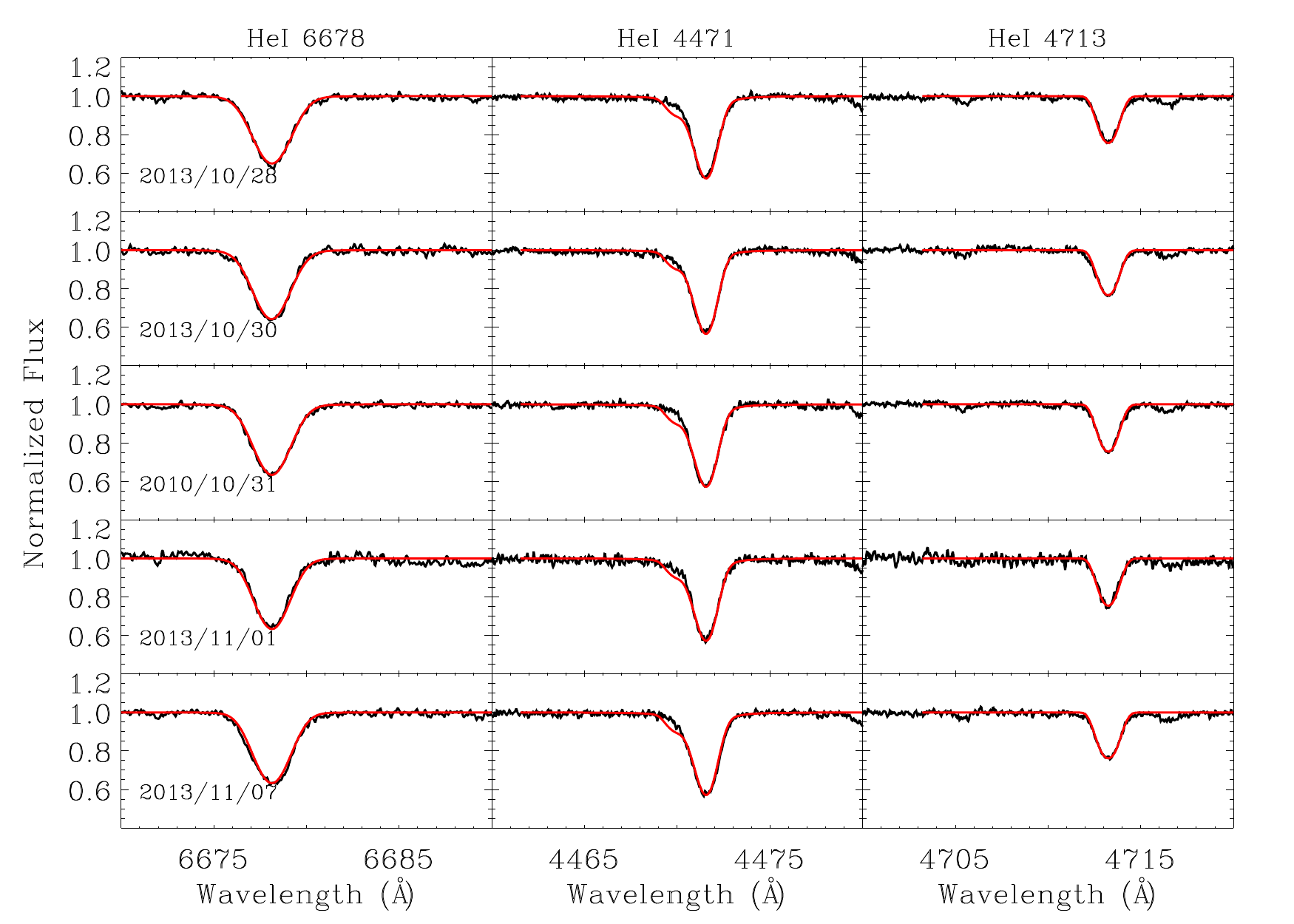}
 \caption{Continued.}
\hspace*{0.5cm}
\label{fig:continued:three}
 \end{center}
\end{figure*}


\subsubsection{Stellar parameters}

On average, we represent the photosphere of the star with  
$T_{\rm{eff}} = 18\,800$\,K, with variations ranging from 18\,570 K to 
19\,100 K (see Tables \ref{tab2:results} and \ref{tab1:results}).
We find a mean value for $\log\,g$ of 2.43 with a scatter of 
$\pm$0.14 dex, based on the fits of the H$\gamma$ line profile.
The same trend is obtained from the Ond\v{r}ejov spectra 
based on the analysis of the H$\alpha$ line alone. Although these latter 
$\log\,g$ values can be less accurate, they range from 2.3 to 2.5 dex, but 
their average is the same ($2.43\pm 0.1$ dex) as the mean value determined 
from the Arizona spectra. Both results predict marginally higher $\log\,g$ 
values than those previously reported in the literature ($\log\,g = 2.2\pm 
0.1$), which are typically based on only one or two measurements, but
the agreement within the error bars is fairly good. 

From the 15 Arizona spectra we obtain an average radius of $57\,R_\odot$
with a statistical dispersion of $\pm 1\,R_\odot$. This radius is close to the 
initial input value of $61\,R_\odot$. We noticed that changes in $\Delta R_{*}$
by more than $1\,R_\odot$ result in appreciable changes in the computed 
profiles. Moreover, several of the 43 H$\alpha$ profiles in the Ond\v{r}ejov
spectra (spread over four years) could not be modeled with a fixed radius of
$57\pm 1\,R_\odot$. Instead, best-fit models were obtained for values
ranging from $52\,R_\odot$ to $65\,R_\odot$. This suggests that the star
either has a variable wind opacity or undergoes real radial changes, for example, 
via radial pulsations.

Using the mean value for R$_\star$ and the surface gravity
corrected for rotation, $\log\,g_{\rm true} = 2.44\,$dex, we determine a  
stellar mass of $34\pm 4$\,M$_\odot$. The mean stellar luminosity is 
found to be $\log L/$L$_\odot = (5.57\pm 0.03)$, which agrees 
with values derived from observations. 

\begin{table*}[ht]
\begin{center}
\caption{Stellar and wind parameters obtained from line profile fits of
 observations taken in Arizona in 2013. 
Listed are the date of observation, the Julian date (JD), the derived effective temperature ($T_{\rm{eff}}$) and 
$\log\,g$, the $\beta$ exponent of the velocity law, the mass-loss rate ($\dot{M}$), the terminal velocity ($v_\infty$), 
the macroturbulent velocity in H$\alpha$ ($v_{\rm{macro}}$(H$\alpha$)), the stellar radius ($R_\star$), and the 
luminosity ($L$).
Mass-loss rates are for smooth wind models. 
\label{tab2:results}}
\begin{tabular}{cccccccccc}
\hline
\hline
Date     & JD & $T_{\rm {eff}}$ & $\log\,g$ & $\beta$  &           $\dot{M}$           & $v_{\infty}$ & $v_{\rm{macro}}$(H$\alpha$) & $R_{\star}$ & $\log\,L/L_\odot$ \\
dd/mm/yr &    &      [K]       &     [cgs]      &          & [$10^{-6}$\,$M_{\odot}$\,yr$^{-1}$] &     [km\,s$^{-1}$]     &   [km\,s$^{-1}$]     & [$R_{\odot}$] & \\
\hline
17/10/2013   &  2\,456\,582.72    &    19\,000$\pm400$   &
2.50$\pm0.1$   &   2.0   &    0.295$\pm0.01$      &     310$\pm30$    &
50$\pm5$ &  58$\pm1$    &  5.60$\pm0.03$ \\
18/10/2013   &  2\,456\,583.70    &    19\,000$\pm400$   &
2.50$\pm0.1$   &   2.0   &    0.295$\pm0.01$      &     310$\pm30$    &
20$\pm5$ &  58$\pm1$    &  5.60$\pm0.03$ \\
19/10/2013   &  2\,456\,584.62    &    19\,000$\pm400$   &
2.50$\pm0.1$   &   2.0   &    0.310$\pm0.01$      &     320$\pm30$    &
35$\pm5$ &  58$\pm1$    &  5.60$\pm0.03$ \\
20/10/2013   &  2\,456\,585.67    &    19\,000$\pm400$   &
2.47$\pm0.1$   &   2.0   &    0.320$\pm0.01$      &     320$\pm30$    &
50$\pm5$ &  57$\pm1$    &  5.59$\pm0.03$ \\
21/10/2013   &  2\,456\,586.73    &    19\,100$\pm400$   &
2.50$\pm0.1$   &   2.0   &    0.280$\pm0.01$      &     300$\pm30$    &
60$\pm5$ &  57$\pm1$    &  5.60$\pm0.03$ \\
22/10/2013   &  2\,456\,587.70    &    19\,000$\pm400$   &
2.40$\pm0.1$   &   2.0   &    0.285$\pm0.01$      &     310$\pm30$    &
60$\pm5$ &  58$\pm1$    &  5.60$\pm0.03$ \\
23/10/2013   &  2\,456\,588.59    &    19\,000$\pm400$   &
2.40$\pm0.1$   &   2.0   &    0.285$\pm0.01$      &     350$\pm35$    &
60$\pm5$ &  58$\pm1$    &  5.57$\pm0.03$ \\
25/10/2013   &  2\,456\,590.59    &    18\,650$\pm400$   &
2.40$\pm0.1$   &   2.0   &    0.190$\pm0.01$      &     300$\pm30$    &
40$\pm5$ &  58$\pm1$    &  5.57$\pm0.03$ \\
26/10/2013   &  2\,456\,591.59    &    18\,600$\pm400$   &
2.40$\pm0.1$   &   2.0   &    0.180$\pm0.01$      &     270$\pm30$    &
50$\pm5$ &  57$\pm1$    &  5.55$\pm0.03$\\
27/10/2013   &  2\,456\,592.59    &    18\,570$\pm400$   &
2.40$\pm0.1$   &   2.0   &    0.170$\pm0.01$      &     270$\pm30$    &
50$\pm5$ &  56$\pm1$    &  5.53$\pm0.03$ \\
28/10/2013   &  2\,456\,593.60    &    18\,570$\pm400$   &
2.40$\pm0.1$   &   2.0   &    0.170$\pm0.01$      &     270$\pm30$    &
40$\pm5$ &  56$\pm1$    &  5.53$\pm0.03$ \\
30/10/2013   &  2\,456\,595.71    &    19\,000$\pm400$   &
2.36$\pm0.1$   &   2.2   &    0.284$\pm0.01$      &     330$\pm30$    &
40$\pm5$ &  57$\pm1$    &  5.58$\pm0.03$ \\
31/10/2013   &  2\,456\,596.71    &    18\,600$\pm400$   &
2.40$\pm0.1$   &   2.0   &    0.180$\pm0.01$      &     270$\pm30$    &
60$\pm5$ &  57$\pm1$    &  5.55$\pm0.03$ \\
01/11/2013   &  2\,456\,597.57    &    18\,600$\pm400$   &
2.40$\pm0.1$   &   2.0   &    0.175$\pm0.01$      &     260$\pm30$    &
60$\pm5$ &  57$\pm1$    &  5.55$\pm0.03$ \\
07/11/2013   &  2\,456\,603.59    &    18\,570$\pm400$   &
2.40$\pm0.1$   &   2.0   &    0.170$\pm0.01$      &     270$\pm30$    &
40$\pm5$ &  56$\pm1$    &  5.53$\pm0.03$ \\
\hline
\end{tabular}
\end{center}
\end{table*}

\begin{table*}[ht]
\begin{center}
\caption{
As Table\,\ref{tab2:results}, but for the  
observations taken in Ond\v{r}ejov between 2009 and 2013. 
\label{tab1:results}}
\begin{tabular}{ccccccccccc}
\hline
\hline
Date     & JD & $T_{\rm {eff}}$ & $\log\,g$ & $\beta$  &           $\dot{M}$           & $v_{\infty}$ & $v_{\rm{macro}}$(H$\alpha$) & $R_{\star}$ & $\log\,L/L_\odot$ \\
dd/mm/yr &    &      [K]       &     [cgs]      &          & [$10^{-6}$\,$M_{\odot}$\,yr$^{-1}$] &     [km\,s$^{-1}$]     &    [km\,s$^{-1}$]     & [$R_{\odot}$] & \\
\hline
15/08/2009   &  2455059.44    &    19000$\pm1000$   &  2.40$\pm0.1$   &
2.0  &    0.290$\pm0.01$      &    310$\pm30$    &   35$\pm5$    &
58$\pm2$    &  5.60$\pm$0.03 \\
18/08/2009   &  2455062.43    &    19000$\pm1000$   &  2.43$\pm0.1$   &
2.0  &    0.295$\pm0.01$     &     340$\pm35$   &   50$\pm5$    & 54$\pm2$
&  5.54$\pm$0.03 \\
23/08/2009   &  2455067.49    &    19000$\pm1000$   &  2.43$\pm0.1$   &
2.0  &    0.340$\pm0.01$     &     320$\pm30$    &   30$\pm5$    &
55$\pm2$    &  5.55$\pm$0.03 \\
\hline
16/07/2010   &  2455394.41    &    18700$\pm1000$   &  2.35$\pm0.1$   &
2.0  &    0.240$\pm0.01$      &     250$\pm25$    &   14$\pm5$    &
63$\pm2$    &  5.64$\pm$0.03 \\
31A/07/2010  &  2455409.37    &    18700$\pm1000$   &  2.50$\pm0.1$   &
2.0  &    0.195$\pm0.01$      &     250$\pm25$    &   35$\pm5$    &
57$\pm2$    &  5.56$\pm$0.03 \\
31B/07/2010  &  2455409.53    &    18700$\pm1000$   &  2.50$\pm0.1$   &
2.0  &    0.195$\pm0.01$      &     250$\pm25$    &   40$\pm5$   &
57$\pm2$    &  5.56$\pm$0.03 \\
01A/08/2010  &  2455410.35    &    18700$\pm1000$   &  2.50$\pm0.1$   &
2.0  &    0.195$\pm0.02$      &     250$\pm50$    &   80$\pm10$    &
57$\pm3$    &  5.56$\pm$0.06 \\
01B/08/2010  &  2455410.41    &    18700$\pm1000$   &  2.50$\pm0.1$   &
2.0  &    0.195$\pm0.02$      &     250$\pm50$    &   90$\pm10$    &
57$\pm3$    &  5.56$\pm$0.06 \\
01C/08/2010  &  2455410.45    &    18700$\pm1000$   &  2.50$\pm0.1$   &
2.0  &    0.195$\pm0.02$      &     250$\pm50$    &   100$\pm10$   &
57$\pm3$    &  5.56$\pm$0.06 \\
04A/09/2010  &  2455444.38    &    18650$\pm1000$   &  2.50$\pm0.1$   &
2.0  &    0.175$\pm0.01$      &     270$\pm30$    &   20$\pm5$    &
56$\pm2$    &  5.54$\pm$0.03 \\
04B/09/2010  &  2455444.41    &    18650$\pm1000$   &  2.50$\pm0.1$   &
2.0  &    0.175$\pm0.01$      &     270$\pm30$    &   30$\pm5$    &
56$\pm2$    &  5.54$\pm$0.03 \\
04C/09/2010  &  2455444.45    &    18650$\pm1000$   &  2.50$\pm0.1$   &
2.0  &    0.175$\pm0.01$      &     270$\pm30$    &   40$\pm5$    &
56$\pm2$    &  5.54$\pm$0.03 \\
05/09/2010   &  2455445.31    &    19000$\pm1000$   &  2.41$\pm0.1$   &
2.2  &    0.280$\pm0.01$      &     310$\pm30$    &   50$\pm5$    &
55$\pm2$    &  5.55 $\pm$0.03\\
06/09/2010   &  2455446.52    &    19000$\pm1000$   &  2.41$\pm0.1$   &
2.2  &    0.280$\pm0.01$      &     310$\pm30$    &   60$\pm5$    &
55$\pm2$    &  5.55$\pm$0.03 \\
11/09/2010   &  2455451.27    &    18600$\pm1000$   &  2.50$\pm0.1$   &
2.0  &    0.195$\pm0.01$      &     250$\pm25$    &   120$\pm10$   &
57$\pm2$    &  5.55$\pm$0.03 \\
21/09/2010   &  2455461.34    &    18600$\pm1000$   &  2.40$\pm0.1$   &
2.0  &    0.180$\pm0.02$      &     250$\pm50$    &   30$\pm5$    &
58$\pm3$    &  5.56$\pm$0.06 \\
22/09/2010   &  2455462.32    &    18600$\pm1000$   &  2.48$\pm0.1$   &
2.0  &    0.153$\pm0.02$      &     235$\pm50$    &   55$\pm5$    &
58$\pm3$    &  5.64$\pm$0.06 \\
23/09/2010   &  2455463.38    &    18700$\pm1000$   &  2.42$\pm0.1$   &
2.0  &    0.236$\pm0.01$      &     260$\pm25$    &   50$\pm5$    &
63$\pm2$    &  5.56$\pm$0.03 \\
24/09/2010   &  2455464.38    &    19000$\pm1000$   &  2.45$\pm0.1$   &
2.0  &    0.305$\pm0.01$      &     330$\pm35$    &   35$\pm5$    &
54$\pm2$    &  5.54$\pm$0.03 \\
\hline
06/03/2011   &  2455627.64    &    19000$\pm1000$   &  2.43$\pm0.1$   &
2.0  &    0.335$\pm0.01$      &     320$\pm30$   &   30$\pm5$    &
55$\pm2$    &  5.55$\pm$0.03 \\
08/03/2011   &  2455629.60    &    19000$\pm1000$   &  2.43$\pm0.1$   &
2.0  &    0.360$\pm0.01$      &     320$\pm30$    &   20$\pm5$    &
55$\pm2$    &  5.55$\pm$0.03 \\
19/03/2011   &  2455640.63    &    19000$\pm1000$   &  2.43$\pm0.1$   &
2.0  &    0.370$\pm0.01$      &     320$\pm30$    &   100$\pm10$   &
55$\pm2$    &  5.57$\pm$0.03 \\
06/09/2011   &  2455811.40    &    18700$\pm1000$   &  2.46$\pm0.1$   &
2.0  &    0.265$\pm0.01$      &     300$\pm30$    &   30$\pm5$    &
58$\pm2$    &  5.58$\pm$0.03 \\
25/09/2011   &  2455830.41    &    18900$\pm1000$   &  2.45$\pm0.1$   &
2.0  &    0.400$\pm0.01$      &     700$\pm70$    &   40$\pm5$    &
55$\pm2$    &  5.55$\pm$0.03 \\
16/10/2011   &  2455851.38    &    19100$\pm1000$   &  2.50$\pm0.1$   &
2.0  &    0.238$\pm0.01$      &     350$\pm35$    &   50$\pm5$    &
53$\pm2$    &  5.53$\pm$0.03 \\
\hline
26/02/2012   &  2455984.63    &    19000$\pm1000$   &  2.50$\pm0.1$   &
2.0       &    0.460$\pm0.01$      &     600$\pm60$    &  30$\pm5$    &
52$\pm2$    &  5.50$\pm$0.03 \\
30/09/2012   &  2456201.23    &    18600$\pm1000$   &  2.30$\pm0.1$   &
2.0       &    0.210$\pm0.01$     &     230$\pm25$    &   80$\pm10$    &
65$\pm2$    &  5.67$\pm$0.03 \\
03/10/2012   &  2456204.26    &    18600$\pm1000$   &  2.30$\pm0.1$   &
2.0       &    0.200$\pm0.02$      &     230$\pm50$    &   30$\pm5$    &
65$\pm3$    & 5.67$\pm$0.06 \\
14/10/2012   &  2456215.28    &    19000$\pm1000$   &  2.43$\pm0.1$   &
2.0       &    0.293$\pm0.01$      &     340$\pm35$    &   30$\pm5$    &
54$\pm2$    &  5.54$\pm$0.03 \\
14/11/2012   &  2456246.24    &    19100$\pm1000$   &  2.46$\pm0.1$   &
2.0       &    0.200$\pm0.02$      &     450$\pm90$    &   70$\pm5$    &
58$\pm3$    &  5.61$\pm$0.06 \\
\hline
16/03/2013   &  2456368.52    &    19000$\pm1000$   &  2.45$\pm0.1$   &
2.0       &    0.300$\pm0.01$      &     300$\pm30$    &   40$\pm5$    &
52$\pm2$    &  5.51$\pm$0.03 \\
21/04/2013   &  2456404.47    &    19000$\pm1000$   &  2.45$\pm0.1$   &
2.0       &    0.305$\pm0.01$      &     330$\pm35$    &   40$\pm5$    &
54$\pm2$    &  5.54$\pm$0.03 \\
25/04/2013   &  2456408.47    &    19000$\pm1000$   &  2.45$\pm0.1$   &
2.0       &    0.305$\pm0.01$      &     330$\pm35$    &   30$\pm5$    &
54$\pm2$    &  5.54$\pm$0.03 \\
19/07/2013   &  2456493.37    &    19000$\pm1000$   &  2.40$\pm0.1$   &
2.0       &    0.295$\pm0.01$      &     310$\pm30$    &   35$\pm5$    &
58$\pm2$    &  5.60$\pm$0.03 \\
20/07/2013   &  2456494.35    &    18700$\pm1000$   &  2.35$\pm0.1$   &
2.0       &    0.200$\pm0.01$      &     250$\pm25$    &   50$\pm5$    &
63$\pm2$    &  5.65$\pm$0.03 \\
21/07/2013   &  2456495.33    &    18700$\pm1000$   &  2.35$\pm0.1$   &
2.0       &    0.200$\pm0.01$      &     250$\pm25$    &   65$\pm5$    &
63$\pm2$    &  5.65$\pm$0.03 \\
22/07/2013   &  2456496.35    &    18700$\pm1000$   &  2.35$\pm0.1$   &
2.0       &    0.200$\pm0.01$      &     250$\pm25$    &   110$\pm10$   &
63$\pm2$    &  5.65$\pm$0.03 \\
23/07/2013   &  2456497.52    &    18700$\pm1000$   &  2.35$\pm0.1$   &
2.0       &    0.200$\pm0.01$      &     250$\pm25$    &   80$\pm10$    &
63$\pm2$    &  5.65$\pm$0.03 \\
25/07/2013   &  2456499.55    &    18700$\pm1000$   &  2.35$\pm0.1$   &
2.0       &    0.200$\pm0.02$      &     250$\pm50$    &   150$\pm15$   &
63$\pm3$    &  5.65$\pm$0.06 \\
30/07/2013   &  2456504.35    &    18700$\pm1000$   &  2.35$\pm0.1$   &
2.0       &    0.200$\pm0.01$      &     250$\pm25$    &   55$\pm5$    &
63$\pm2$    &  5.65$\pm$0.03 \\
01/08/2013   &  2456506.57    &    18700$\pm1000$   &  2.40$\pm0.1$   &
1.6       &    0.215$\pm0.01$      &      250$\pm25$    &   40$\pm5$    &
60$\pm2$    & 5.60$\pm0.03$ \\
02/08/2013   &  2456507.34    &    19000$\pm1000$   &  2.43$\pm0.1$   &
2.0       &    0.295$\pm0.01$      &     340$\pm35$    &   80$\pm10$    &
54$\pm2$    &  5.54$\pm0.03$ \\
03/08/2013   &  2456508.34    &    19000$\pm1000$   &  2.43$\pm0.1$   &
2.0       &    0.295$\pm0.01$      &     340$\pm35$    &   70$\pm5$    &
54$\pm2$    &  5.54$\pm0.03$ \\
\hline
\end{tabular}
\tablefoot{$T_{\rm eff}$ values are obtained from simultaneous fits to the H and He lines. The table lists only those models for which it was possible to derive the model parameters. Modeled and non-modeled H$\alpha$ profiles are shown in Figs.\,\ref{halpha2009}-\ref{halpha2012}.}
\end{center}
\end{table*}

\subsubsection{Wind variability}
\label{wind}

To gain insight into the variability of the stellar wind of 
55\,Cyg,
we modeled the H$\alpha$ line profile over the observing period
of $\text{about}$  five years.

We fit the observations assuming a smooth (unclumped) wind structure 
represented by a $\beta$-velocity law, with $\beta = 2.0$, except 
in rare cases, where the value of $\beta$ was slightly adjusted (1.6 and 2.2). On 
average, the derived $\dot{M}$ from the Ond\v{r}ejov data is 
$\sim 2.5\times 10^{-7}$\,M$_\odot$\,yr$^{-1}$, with fluctuations of a 
factor of three between the minimum value of 
$1.5\times 10^{-7}$\,M$_\odot$\,yr$^{-1}$ and the maximum value of 
$4.6\times 10^{-7}$\,M$_\odot$\,yr$^{-1}$. 
Our results agree on average with the homogeneous wind parameters reported by 
\citetads{2006A&A...446..279C}: $\beta = 2$ and  
$\dot{M} = 2.3\times 10^{-7}$\,M$_\odot$\,yr$^{-1}$, and our values for the 
mass-loss fall between the extreme ones obtained for unclumped winds 
in the literature (see Sect.\,\ref{sec:star}). In addition, the 
variation in the terminal velocity (ranging typically from 230\,km\,s$^{-1}$ 
to 350\,km\,s$^{-1}$) presents a positive correlation with the mass-loss rate. 
There are two exceptions with substantially higher values of 600\,km\,s$^{-1}$ 
and 700\,km\,s$^{-1}$, corresponding to the two occasions of highest mass loss of 
$4.6\times 10^{-7}$\,M$_\odot$\,yr$^{-1}$ and 
$4.0\times 10^{-7}$\,M$_\odot$\,yr$^{-1}$. 
The corresponding observations were taken on 2011 September 25 and 2012 February 
26, when a mass eruptive event probably occurred.
The terminal velocities found are higher than the 200\,km\,s$^{-1}$ reported by 
\citetads{Markova2008} and lower than the 470\,km\,s$^{-1}$ adopted by 
\citetads{2006A&A...446..279C}. 

The average mass-loss rate obtained 
(Ond\v{r}ejov observations) is consistent with the average $\dot{M}$ value 
estimated from the high-resolution spectra taken in Arizona during October and 
November, 2013. Furthermore, the Arizona spectra show changes in the intensity 
of the H$\alpha$ line (as shown in Fig. \ref{fig:continued:one}) and 
variations in the mass-loss rate by a factor of 1.9.

\subsubsection{Turbulent and projected rotation velocities}
\label{vel}

In addition to the stellar and wind parameters discussed so far,
three more parameters were used to 
match the observed line profiles. These are microturbulence, macroturbulence, 
and projected rotation velocity. Microturbulent velocities refer to 
atmospheric motions on scales below the photon mean free path, and there is 
evidence that these motions could result from subsurface convection caused by 
the iron opacity peak \citepads{2009A&A...499..279C}. 
The values found for BSGs range between 
5 km\,s$^{-1}$ and 15 km\,s$^{-1}$ with a tendency to decrease with decreasing 
temperature \citepads[e.g.,][]{Markova2008}. Fits to the helium
absorption lines give best results for $10\pm 2$\,km\,s$^{-1}$, 
except for \ion{He}{i}\,$\lambda$\,6678, which requires a higher value 
($17\pm 2$\,km\,s$^{-1}$). Similar values are observed in \ion{Si}{ii} and \ion{Si}{iii} lines, with $12\pm 2$\,km\,s$^{-1}$ and $7\pm 2$\,km\,s$^{-1}$, respectively.
The line profiles of H$\alpha$ also require a
higher microturbulent velocity to achieve 
satisfactory fits. Here, a value of $\sim 50\pm 5$\,km\,s$^{-1}$ is needed, which can
be explained by the fact that H$\alpha$ is formed in the lower wind where 
supersonic turbulent velocities can occur. Problems 
in fitting H$\alpha$ line profiles are not new, and, for many BSGs, 
values as high as 50 km\,s$^{-1}$ (or turbulent velocities increasing linearly
from the photospheric value to about 50 km\,s$^{-1}$) were reported 
\citepads[e.g.,][]{2006A&A...446..279C, 2008A&A...481..777S}. A possible interpretation
of the greater microturbulent velocity value for \ion{He}{i}\,$\lambda$\,6678 
might be that the line-forming region extends into the base of the wind where 
it is affected by the wind turbulence. We return to this point in 
Sect.\,\ref{dis-rotation}.

After including the microturbulence, the synthetic line profiles must be 
broadened by the combined effect of stellar rotation and macroturbulence.
Considering that $v\sin i$, as a global parameter, should be the same for all 
lines, the variability in the line profiles must be assigned to macroturbulent
broadening. For H and He I lines, the best fit was achieved with $v\sin i = 55\pm 5$ km\,s$^{-1}$ (see also Sect.\,\ref{dis-rotation}), but theoretical and  observed  \ion{Si}{ii} and \ion{Si}{iii} lines match better if $v\sin i$ is taken around $40 \pm 5$ km\,s$^{-1}$ or $45 \pm 5$ km\,s$^{-1}$. We also note that the macroturbulence 
can be quite different in various
lines. For instance, from modeling the Arizona data, we found that 
$v_{\rm macro}$ is relatively small in the lines of 
\ion{He}{i}\,$\lambda$\,4471 (6-7 km\,s$^{-1}$), \ion{He}{i}\,$\lambda$\,4713 
(11-12 km\,s$^{-1}$), and H$\delta$ (14-15 km\,s$^{-1}$), and approximately 
constant throughout the observing period. In H$\gamma$, the changes are larger
(14-30 km\,s$^{-1}$). The situation changes for H$\beta$ and 
\ion{He}{i}\,$\lambda$\,6678, for which much higher and strongly variable 
values are found (25-90 km\,s$^{-1}$ and 20-50 km\,s$^{-1}$, respectively).
Those obtained for H$\alpha$ vary in the range 20-60 km\,s$^{-1}$ (see 
Table \ref{tab2:results}).  The \ion{Si}{ii} and \ion{Si}{iii} lines also show a large dispersion in $v_{\rm macro}$, they range from 10 km\,s$^{-1}$  to 30 km\,s$^{-1}$, and 30 km\,s$^{-1}$ to 50 km\,s$^{-1}$, respectively, with uncertanties of $\pm$ 5 km\,s$^{-1}$. 

The data from Ond\v{r}ejov provide macroturbulent velocities only
for H$\alpha$ and \ion{He}{i}\,$\lambda$\,6678\,\AA. \ Over the entire observing 
period, we obtain values in the range 30-60 km\,s$^{-1}$ for 
\ion{He}{i}\,$\lambda$\,6678\,\AA, \ while for H$\alpha$, occasionally values of more 
than 100 km\,s$^{-1}$ are found (see Table \ref{tab1:results}). 
In most spectra, we find that for both lines the 
contribution of macroturbulence to the line broadening is comparable to or 
higher than that from $v\sin i$. Such high values are commonly found in BSGs 
\citepads[e.g.,][]{Ryans2002, Simon-Diaz2007, Markova2008, 2014A&A...562A..37M, 
2014A&A...562A.135S}, and the values of $v\sin i$ and $v_{\rm macro}$ found 
for 55\,Cyg agree well with those of BSGs of comparable mass and 
effective temperature \citepads[see][]{2014A&A...562A..37M}.

\begin{figure}[t!]
   \centering   
   \includegraphics[width=0.9\hsize]{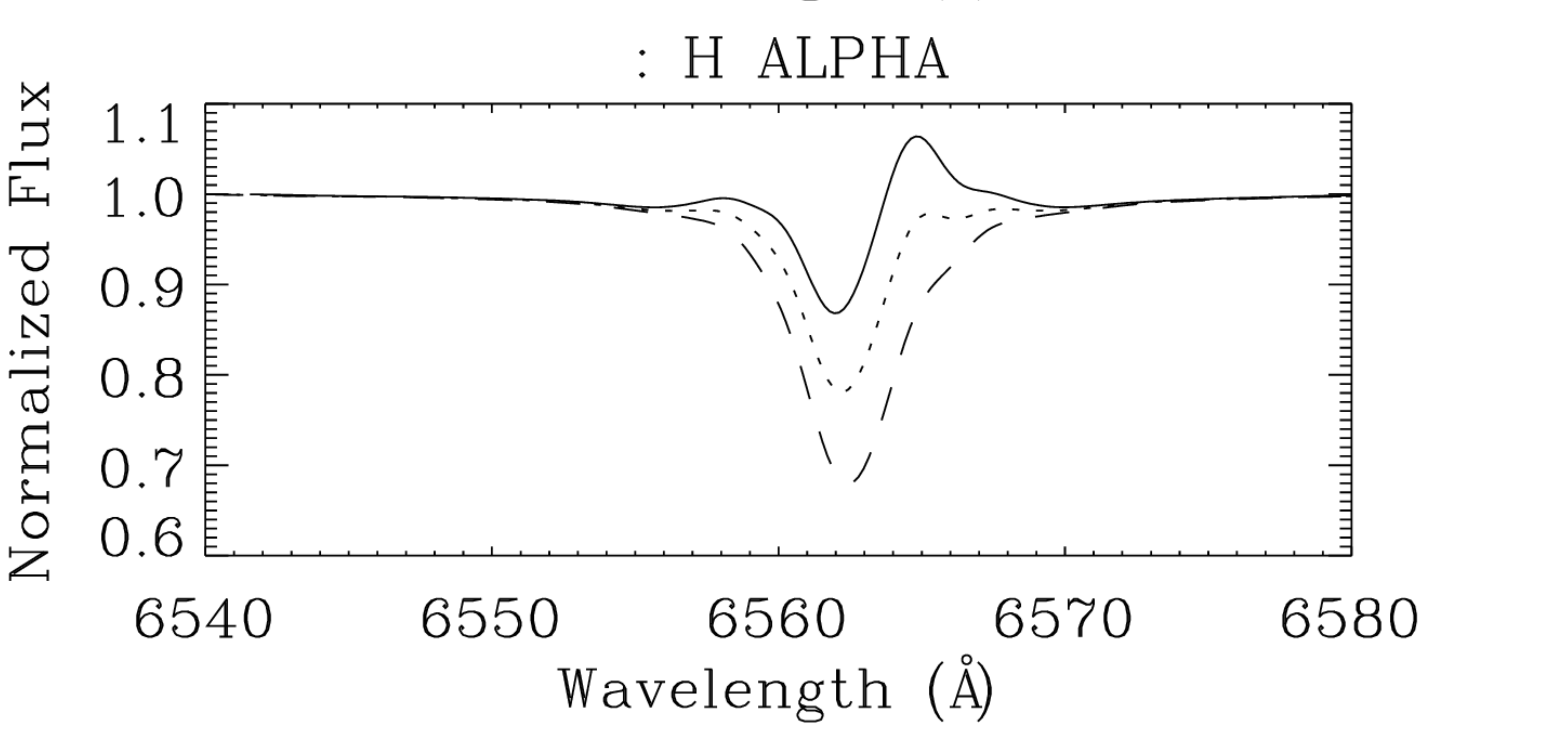}
\includegraphics[width=0.9\hsize]{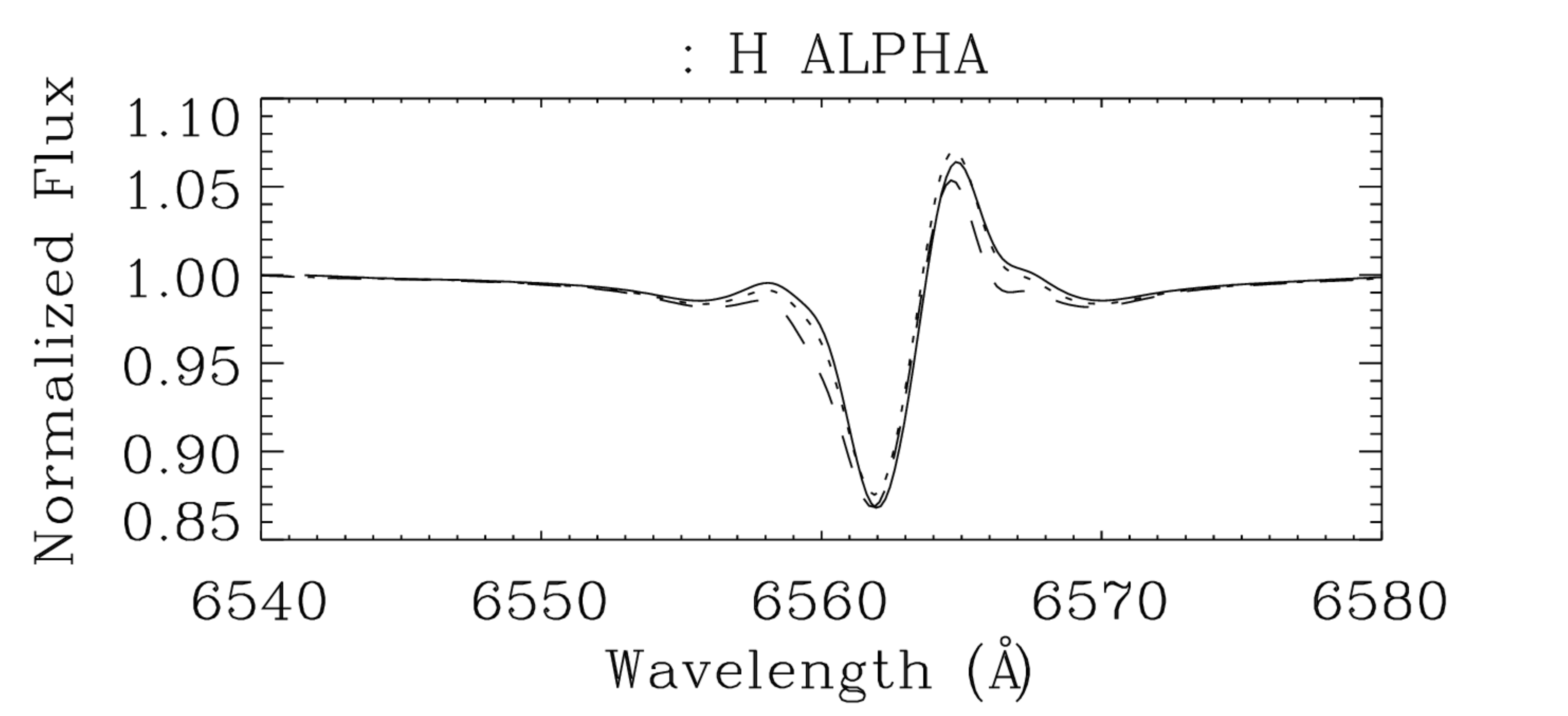}
      \caption {Top panel: Variation of the H$\alpha$ line for different He
      abundances: $N$(He)/$N$(H) = 0.1 (solid line),
      $N$(He)/$N$(H) = 0.2 (dotted line), and $N$(He)/$N$(H) =
      0.4 (dashed line) and the same mass-loss rate. Bottom panel:
      Similar H$\alpha$ line profiles derived using the same $N$(He)/$N$(H)
      ratios, as in the previous plot, but with different mass-loss rates:
      $1.9\times 10^{-7}$\,M$_\odot$\,yr$^{-1}$ (solid line),
      $2.5\times 10^{-7}$\,M$_\odot$\,yr$^{-1}$ (dotted line), and
      $3.5\times 10^{-7}$\,M$_\odot$\,yr$^{-1}$ (dashed line).
         \label{fig:chemical_1}}
  \end{figure}

\begin{figure}[t!]
   \centering   
   \includegraphics[width=\hsize]{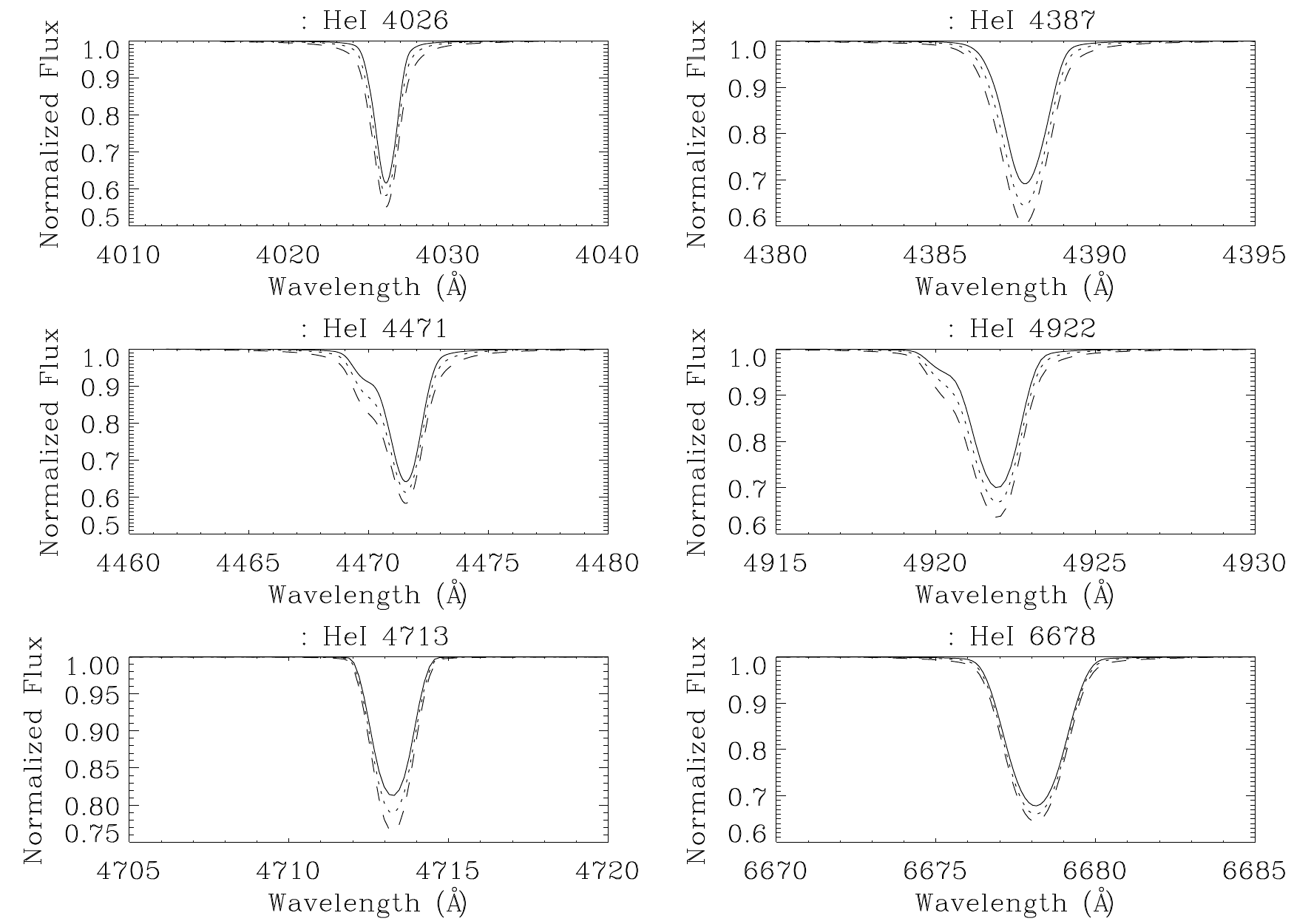}
   \caption{Variation of the \ion{He}{i} lines computed for a rotating star
($V\,\sin\,i$ = 60 km s$^{-1}$) with He contents of $N(He)/N(H)$= 0.1 (solid line),
$N(He)/N(H)$= 0.2 (dotted line) and $N(He)/N(H)$= 0.4 (dashed line). The intensity of
the \ion{He}{i} forbidden components at $\lambda$\,4470 and $\lambda$\,4922 serves 
as an ideal indicator for the He abundance.}
\label{fig:chemical_2}
  \end{figure}

\subsubsection{Atmospheric chemical composition}
\label{chemical}

The H$\alpha$ line is known to be sensitive to variations in the He content, in 
the sense that He-rich wind models produce H$\alpha$ EWs that are a factor of
2-3 lower than those with solar He abundance \citepads{2014A&A...565A..62P}. 
As the mass-loss rate is a function of metallicity and is derived from the strength
of the H$\alpha$ line, we expect a lack of uniqueness in wind models, which are 
based purely on the modeling of H transitions. 

This ambiguity is demonstrated in Fig. \ref{fig:chemical_1}, where we show in 
the top panel the H$\alpha$ line computed for a constant mass-loss rate of $1.9\times 
10^{-7}$\,M$_\odot$\,yr$^{-1}$ and $N$(He)/$N$(H) ratios of 0.1, 0.2, and 0.4.
To compensate for the vanishing H$\alpha$ emission with increasing He abundance, we
must increase the mass-loss rate to $2.5\times 10^{-7}$\,M$_\odot$\,yr$^{-1}$ and
$3.5\times 10^{-7}$\,M$_\odot$\,yr$^{-1}$, respectively. Then all three models
produce approximately identical H$\alpha$ profiles (bottom panel of Fig. \ref{fig:chemical_1}).
Consequently, uncertainties in the chemical composition (here He abundance) 
by a factor of 2--4 may affect the derived mass-loss rate by a factor of 1.3--1.9.

However, a higher He abundance also sensitively alters the strength of both
the permitted and forbidden components of the He lines (see Fig. \ref{fig:chemical_2}). 
When increasing the He abundance for a fixed mass-loss rate, the forbidden components
develop prominent absorption features in the wings of the permitted lines.  
These manifest as pronounced asymmetries in the profiles
of the permitted components, even if the projected rotational velocity is as high
as $\sim 60$\,km\,s$^{-1}$. The presence and strength of the forbidden component
hence serve as useful diagnostics for the He surface abundance of slowly rotating
stars. 

As the He line profiles observed in 55\,Cyg do not show noticeable 
contributions from the forbidden components,  a solar He abundance was found 
to be more appropriate to fit both the H$\alpha$ and He lines.

In relation to the modeling of the \ion{Si}{ii} and \ion{Si}{iii} lines, we 
were able to obtain good fits using solar abundance, but the
question remains of how sensitive the Si lines are to changes
in their content. The influence of non-solar abundance on $T_{\rm eff}$
was discussed in detail by \citet{Lefever2007} and \citetads{Markova2008},
who found that higher abundances result in lower $T_{\rm eff}$ values and vice 
versa. Such a change in temperature ($\Delta T_{\rm eff} > 1\,000$\,K) would
also affect the He lines. However, from our model fits, we do not find a 
discrepancy between the temperature obtained from the He and the Si lines.
Therefore, the assumption of solar abundance for Si (within a range $\pm 
0.15$\,dex) is well justified. All other elements are treated as background 
elements with solar abundance.

\begin{figure}
   \centering   
   \includegraphics[width=\hsize]{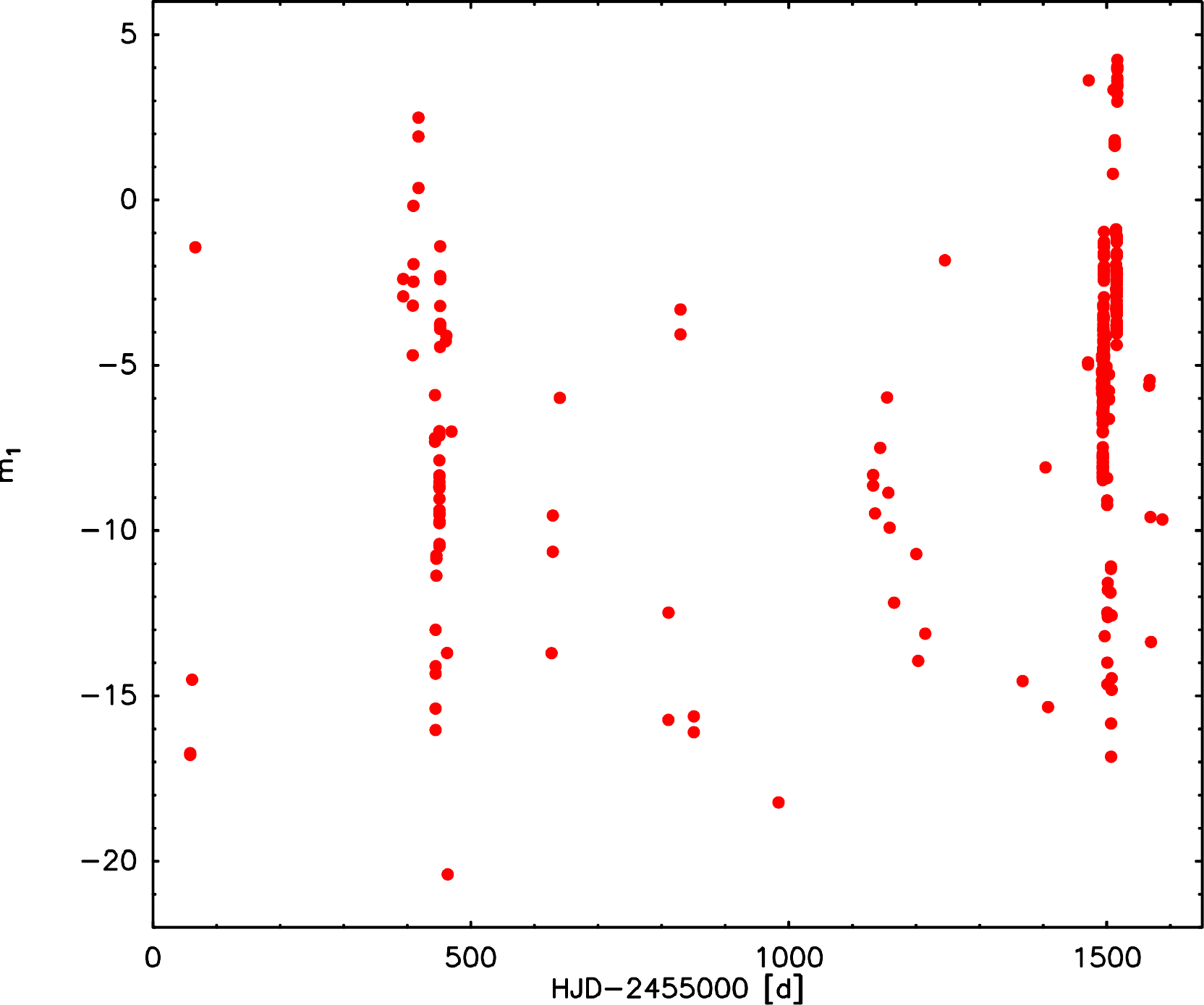}
      \caption{Variation in the first moment (radial velocity, without correction for systemic velocity)
of the \ion{He}{i}\,$\lambda$\,6678 \AA\, line in the period August 2009 to October 2013.}
         \label{fig:complete}
  \end{figure}

\subsection{Searching for pulsations}

As 55\,Cyg resides inside the instability domain for BSGs
\citepads{Saio2006} and $\alpha$ Cygni variables \citepads{Saio2013}, it is 
natural to assume that pulsations might also be at work on the surface of this 
star. The strong variability in the H$\alpha$ line (suggesting variable 
mass-loss) could indicate that the observed changes in the wind might be 
connected to pulsations detectable in photospheric line variations. 
To test this hypothesis, this section is devoted to a detailed variability 
analysis of photospheric line profiles.

\subsubsection{Photospheric line profile variability}

In the absence of long-term photometric light curves, the presence of pulsations 
can also be tested by analyzing the variability of photospheric lines in 
high-quality optical spectra. 
The spectral range observed with the Perek 2m telescope contains several 
photospheric lines (see Fig.\,\ref{fig:spectrum}); the strongest is 
\ion{He}{i}\,$\lambda$6678\,\AA, followed by the deep absorption lines of 
\ion{C}{ii}\,$\lambda\lambda$\,6578,\,6583\,\AA. \ Considerably weaker lines of 
\ion{Si}{ii}\,$\lambda\lambda$\,6347,\,6371\,\AA, \
\ion{Ne}{i}\,$\lambda\lambda$\,6402,\,6507\,\AA, \ and 
\ion{N}{ii}\,$\lambda\lambda$\,6482,\,6611\,\AA \ are present as well. 
Despite the presence of these metal lines, we restricted most of our analysis to 
\ion{He}{i} for two reasons. First, this line is not affected by telluric 
features, contrary to other lines such as H$\alpha$, 
\ion{C}{ii}\,$\lambda\lambda$\,6578,\,6583\,\AA, \
\ion{Si}{ii}\,$\lambda\lambda$\,6347,\,6371\,\AA, \ and 
\ion{N}{ii}\,$\lambda\lambda$\,6482,\,6611\,\AA. \ Any correction of the 
spectra with a telluric spectrum obtained from observing a standard star 
superimposes additional noise to the continuum level of the science spectra. 
Because \ion{He}{i}\,$\lambda$\,6678\,\AA \ falls outside regions of telluric 
pollution, no cleaning is required, and the initial quality of the spectral 
line is preserved. Second, while only the \ion{C}{ii} lines are 
sufficiently strong for an analysis without suffering from an enhanced 
continuum noise level, these lines arise within a broad and shallow emission 
component centered on H$\alpha$ (see Fig.\,\ref{fig:spectrum}). The shape of the line profiles of 
these \ion{C}{ii} lines can thus be (possibly severely) altered by the 
normalization process, imprinting additional errors. 
These metal lines are, therefore, less favorable for the analysis of 
reliable line profile variabilities, but can be used as a secondary 
criterion to prove pulsation activity.

\begin{figure}
   \centering  
   \includegraphics[width=\hsize]{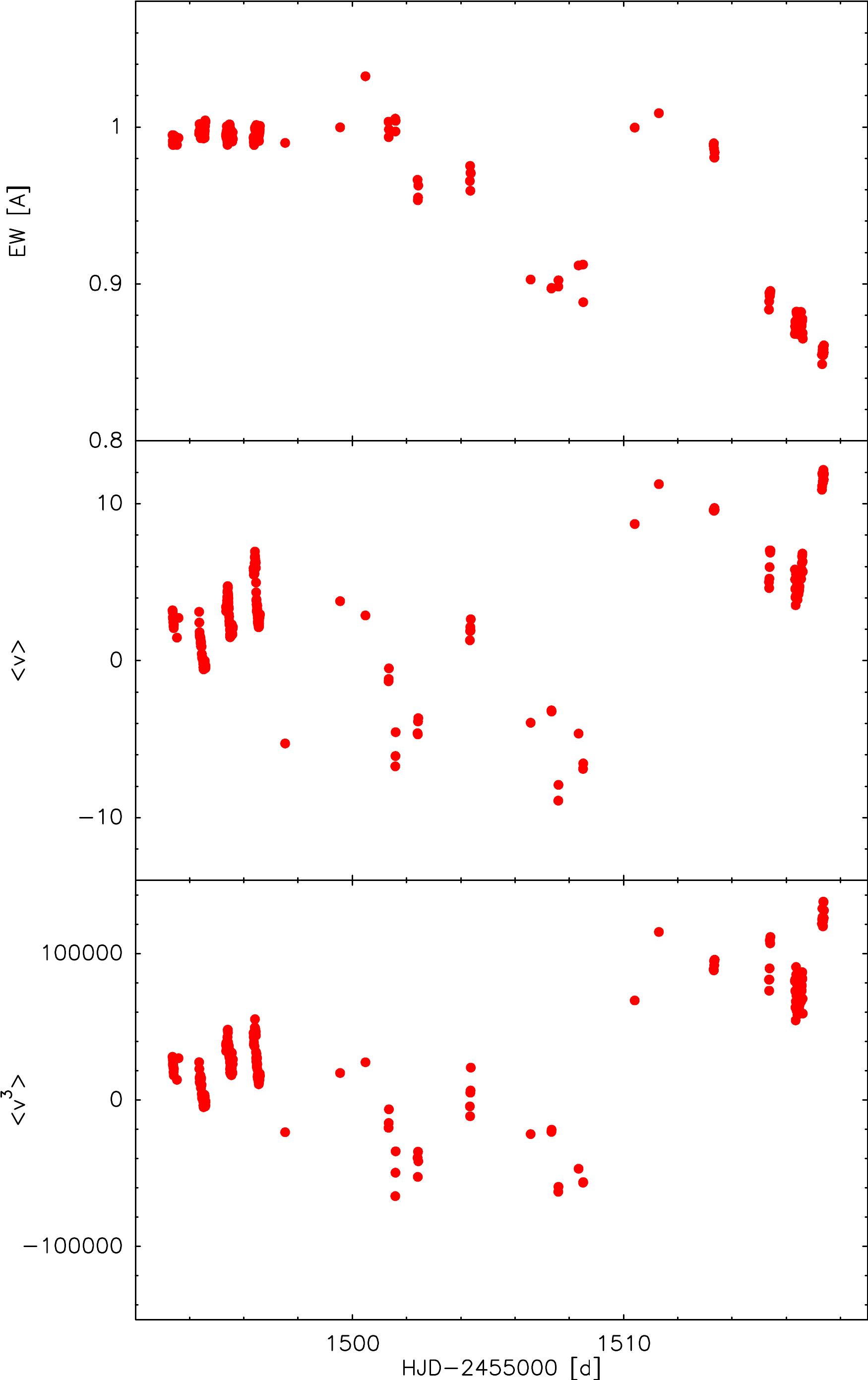}
      \caption{Variation in EW (top), first (radial velocity, middle), and third
        moment (skewness, bottom) of \ion{He}{i}\,$\lambda$\,6678\,\AA\, in the observing period 2013 July 19 to August 12.}
         \label{fig:he_mom}
\end{figure}

\subsubsection{Moment analyses}

For further analysis, we excluded all spectra that have exposure times longer 
than 1800\,s. This leaves us with a total of 325 spectra, most with
exposure times between 400\,s and 900\,s.

To search for line profile variabilities in \ion{He}{i}\,$\lambda$\,6678\,\AA, we applied the
moment method \citepads{1992A&A...266..294A, 1994A&A...288..155N}. This method
is ideal, as it allows distinguishing between line profile variability
caused by stellar spots or pulsations.
We computed the first three moments of \ion{He}{i}\,$\lambda$\,6678\,\AA\, using
the prescription of \citetads{1992A&A...266..294A}. The first moment
corresponds to the radial velocity\footnote{We follow the notation of 
\citetads{2010aste.book.....A}, in which the moments (not normalized and not 
corrected for systemic velocity) are denoted by $m_{i}$, and the normalized,
corrected moments by $\langle\varv^{i}\rangle$.}. This parameter varies between +5\,km\,s$^{-1}$ and
-21\,km\,s$^{-1}$ over the total observing period from August 2009 to
October 2013 (see Fig.\,\ref{fig:complete}). The offset from a symmetric
variation around zero velocity is caused by the star's intrinsic radial
velocity, which we determine from this plot to be $\varv_{\rm rad, star} \simeq
-8$\,km\,s$^{-1}$. We corrected the
data for this stellar motion before we continued with our analysis. 
Hence, the observed maximum amplitude of the radial
velocity variation in the line profile is about $\pm 13$\,km\,s$^{-1}$.

\begin{figure}
   \centering  
   \includegraphics[width=\hsize]{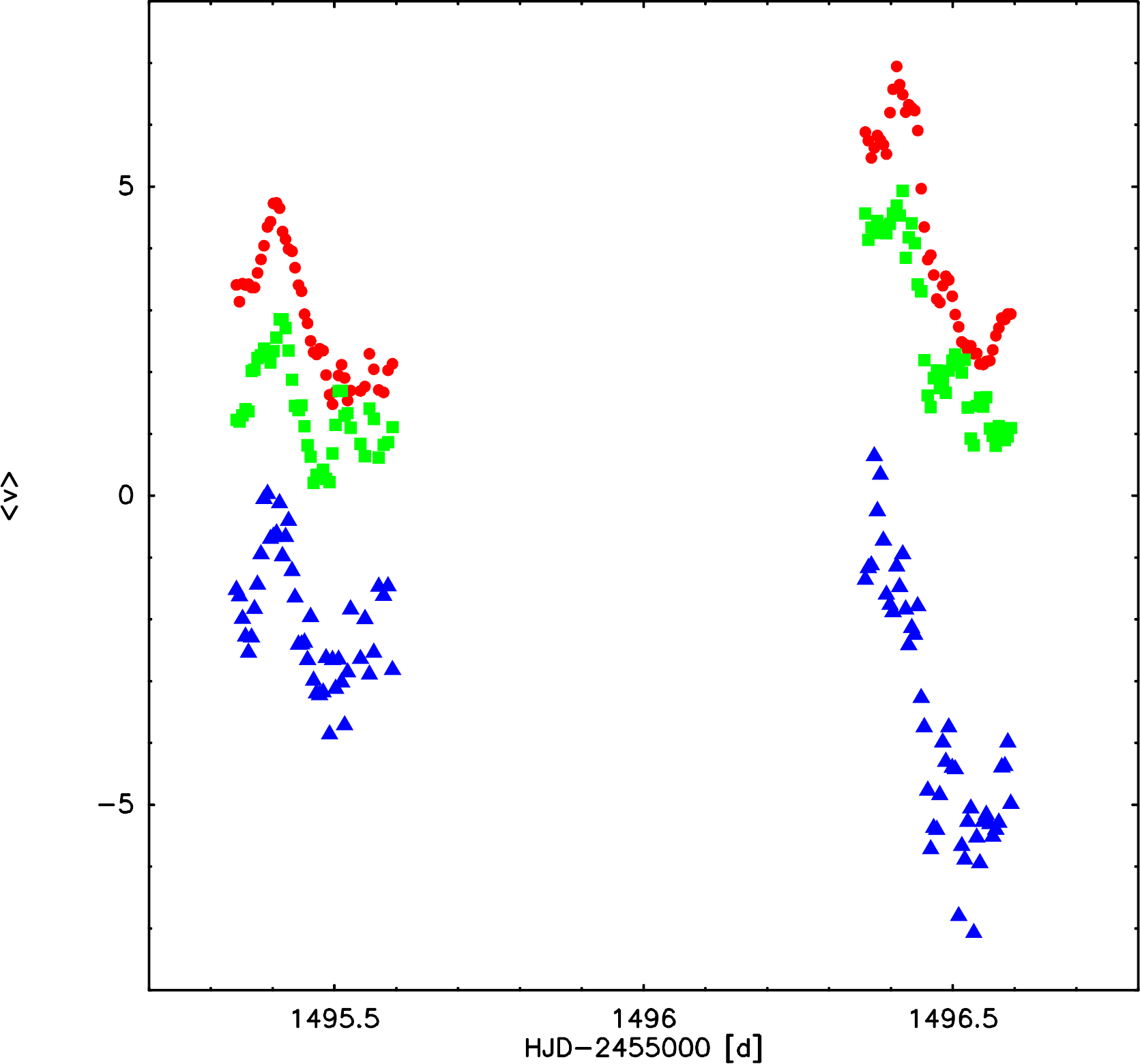}
      \caption{Variation in the first moment (radial velocity)
of \ion{He}{i}\,$\lambda$\,6678\,\AA\, (dots), \ion{C}{ii}\,$\lambda$\,6578\,\AA\, (squares),
and \ion{Si}{ii}\,$\lambda$\,6347\,\AA\, (triangles) during the nights 2013 July 21-22.}
         \label{fig:metall}
  \end{figure}

\begin{table}
\caption{55\,Cyg periodicities}
\label{tab:frequencies}
\centering
\begin{tabular}{lcccccc}
\hline
\hline
No. & $f$ & $df\times 10^{8}$ & $A$ & $dA$ & $\Phi$ & $d\Phi$ \\
    & \multicolumn{2}{c}{(day$^{-1}$)} & \multicolumn{2}{c}{(km\,s$^{-1}$)} & \multicolumn{2}{c}{(rad)} \\
\hline
$f_{1}$  &  0.044366 & 0.03 & 4.39 & 0.03 & 2.43 & 0.01 \\
$f_{2}$  &  0.995570 & 0.03 & 3.31 & 0.04 & 3.71 & 0.02 \\
$f_{3}$  &  0.235716 & 0.07 & 2.54 & 0.03 & 3.44 & 0.06 \\
$f_{4}$  &  0.974983 & 0.17 & 1.42 & 0.03 & 5.40 & 0.09 \\
$f_{5}$  &  1.278593 & 0.18 & 1.37 & 0.03 & 3.35 & 0.10 \\
$f_{6}$  &  2.400132 & 0.17 & 1.54 & 0.02 & 1.21 & 0.11 \\
$f_{7}$  &  0.506733 & 0.22 & 1.30 & 0.07 & 5.55 & 0.13 \\
$f_{8}$  &  4.163120 & 0.26 & 0.87 & 0.01 & 5.78 & 0.17 \\
$f_{9}$  &  6.994100 & 0.21 & 0.95 & 0.01 & 1.76 & 0.14 \\
$f_{10}$ &  3.104664 & 0.18 & 1.18 & 0.02 & 5.29 & 0.10 \\
$f_{11}$ &  0.188665 & 0.42 & 1.09 & 0.04 & 3.43 & 0.26 \\
$f_{12}$ &  1.462843 & 0.40 & 1.06 & 0.03 & 2.77 & 0.25 \\
$f_{13}$ &  0.296804 & 0.24 & 1.11 & 0.03 & 3.31 & 0.14 \\
$f_{14}$ &  0.652575 & 1.21 & 0.79 & 0.03 & 1.24 & 0.77 \\
$f_{15}$ &  1.181013 & 1.55 & 0.55 & 0.04 & 6.03 & 1.00 \\
$f_{16}$ &  0.617903 & 0.30 & 1.00 & 0.02 & 0.04 & 0.17 \\
$f_{17}$ &  2.593881 & 0.45 & 0.75 & 0.02 & 5.46 & 0.28 \\
$f_{18}$ &  9.003491 & 0.52 & 0.54 & 0.01 & 2.09 & 0.34 \\
$f_{19}$ &  3.780456 & 1.10 & 0.51 & 0.02 & 4.94 & 0.79 \\
\hline
\end{tabular}
\tablefoot{Phases are referenced to HJD = 2455059.00\,d.}
\end{table}

\begin{figure}
   \centering  
   \includegraphics[width=\hsize]{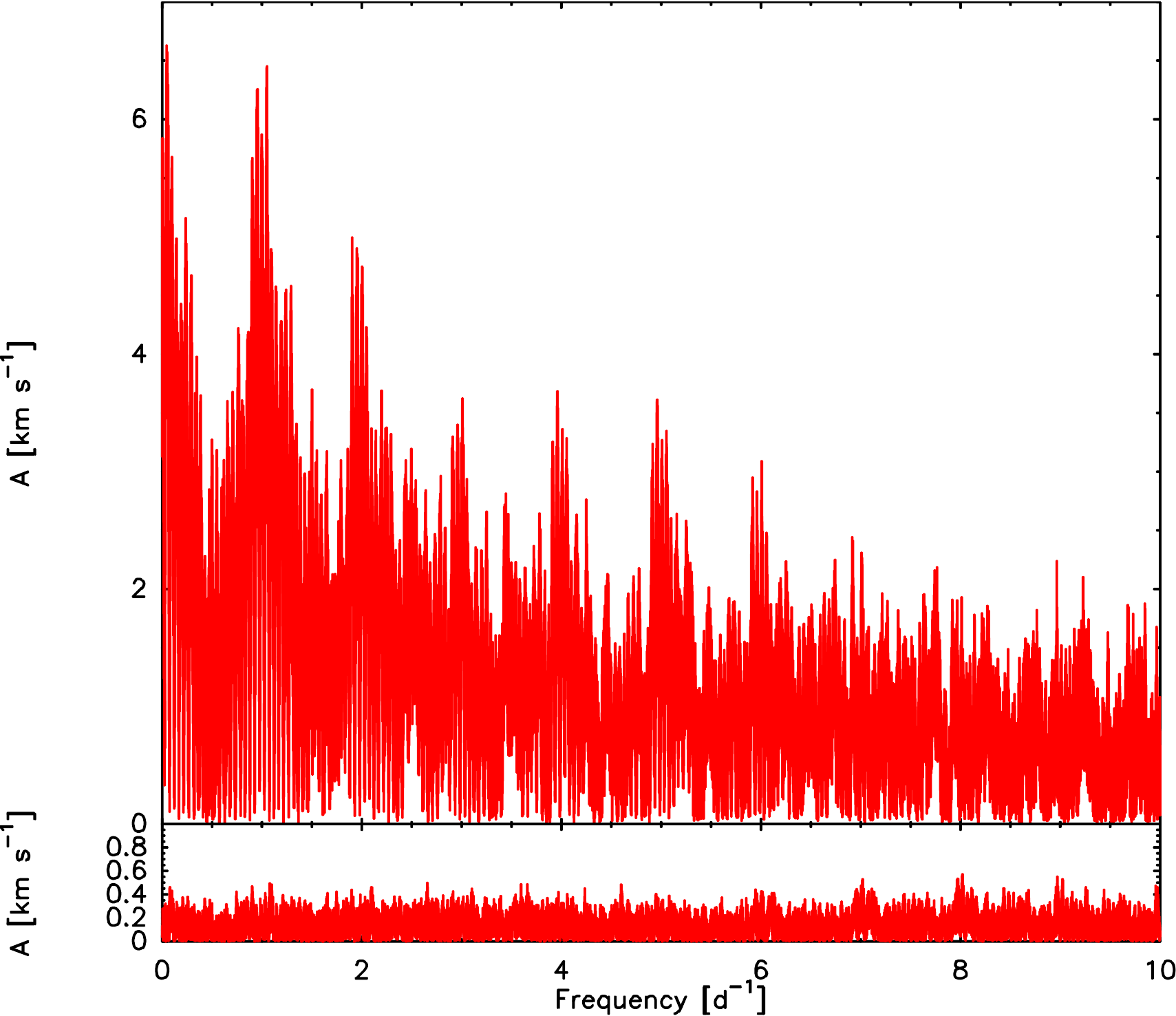}
      \caption{Lomb-Scargle periodogram of the original radial velocity data
 (top) and of the data cleaned with the identified 19 sinusoids (bottom).}
         \label{fig:ampli}
  \end{figure}

Next, we
compared the behavior of the first and third moments. Their time variability is
shown in the middle and bottom panels of Fig.\,\ref{fig:he_mom}. For 
better visibility and clarity, the plot is
limited to a period of 25 days, ranging from 2013 July 19 to August 12. This period
contains time-series within seven nights, four at the beginning and three at the
end. Obviously, the third moment, which describes the skewness (i.e., the
asymmetry) of the line profile displays a strong time variability. 
In addition, the third moment varies in phase with the radial velocity,
which is characteristic for stellar pulsations
\citepads[see, e.g.,][]{1992A&A...266..294A}. Similar behavior is seen in
other seasons.

Additional proof for pulsation activity in the atmosphere of 55\,Cyg is
provided by comparing moments from different elements. Although,
as mentioned above, the other lines in the spectrum have a lower quality, we
computed the moments of the lines \ion{C}{ii}\,$\lambda$\,6578\,\AA\, and \ion{Si}{ii}\,$\lambda$\,6347\,\AA\,, but
restricted to the time-series obtained during the nights 2013 July 21-22, which
have the highest S/N. The results for the first moment of
these lines compared to \ion{He}{i}\,$\lambda$\,6678\,\AA\, are shown in Fig.\,\ref{fig:metall}.
Obviously, all three lines vary identically. Such a behavior
excludes the possibility of stellar spots. If the star has a patchy surface
abundance pattern, as is common in chemically peculiar stars,
for instance, the radial
velocities of the He lines vary typically differently from those of the
metal lines \citepads[e.g.,][]{2001A&A...380..177B, 2004A&A...413..273B,
2006A&A...457.1033L}. Instead, the identical variability of the line profiles
of both He and metals in 55\,Cyg is very typical for pulsations. 

The clear offset of the radial velocity curves of different lines 
could indicate that a velocity gradient exists within the atmosphere of 55\,Cyg.
In addition, the \ion{Si}{ii}\,$\lambda$\,6347\,\AA \ line, which is the 
weakest of these three, displays negative velocities, but the acceleration is
in phase with the other two lines that display positive velocitites. 
Unfortunately, the resolution of our spectra is too low to perform proper mode 
identification, so that no clear statements about the nature of the pulsations 
(radial, non-radial) can be made.  

\begin{figure}
   \centering  
   \includegraphics[width=\hsize,angle=0]{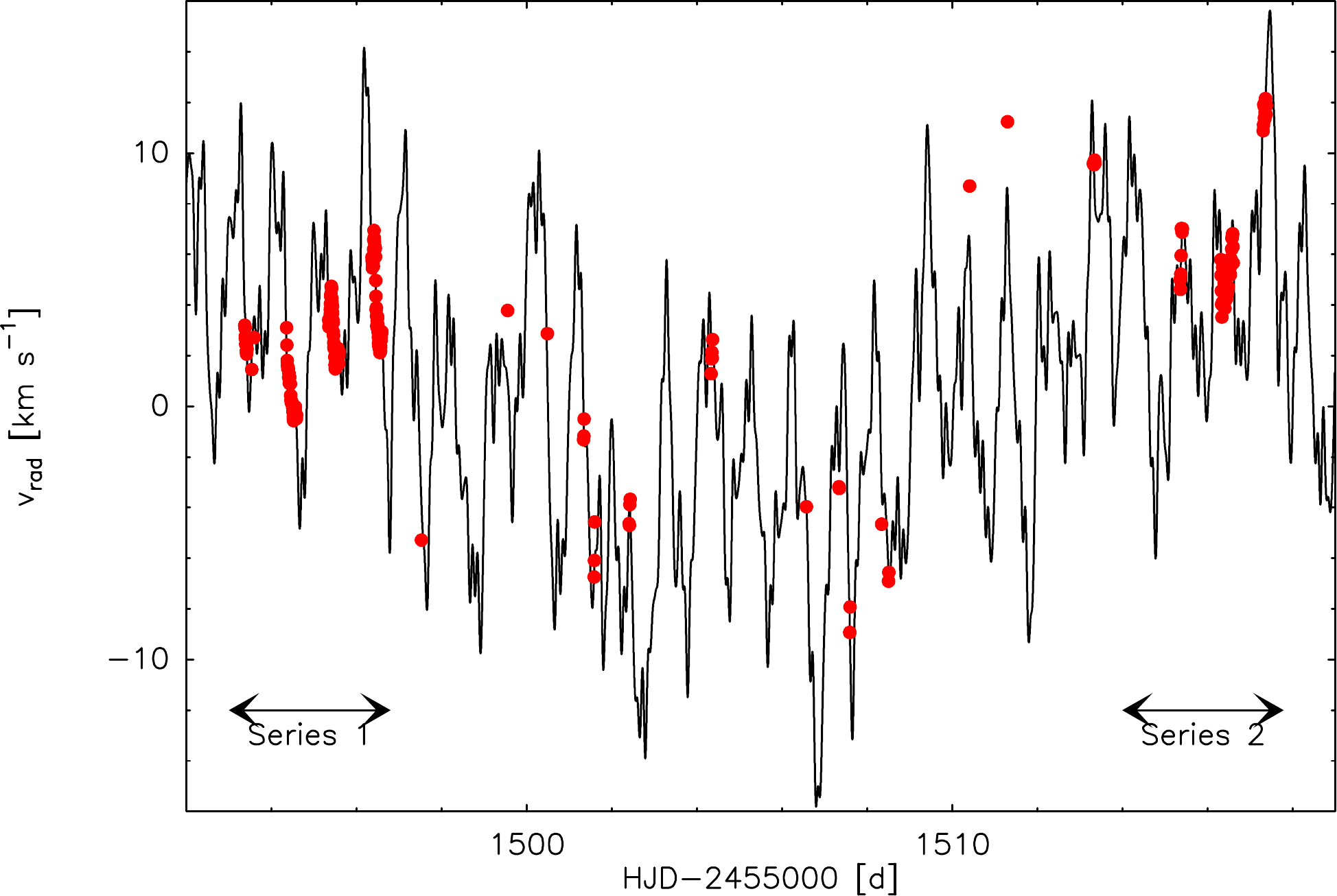}
   \includegraphics[width=\hsize,angle=0]{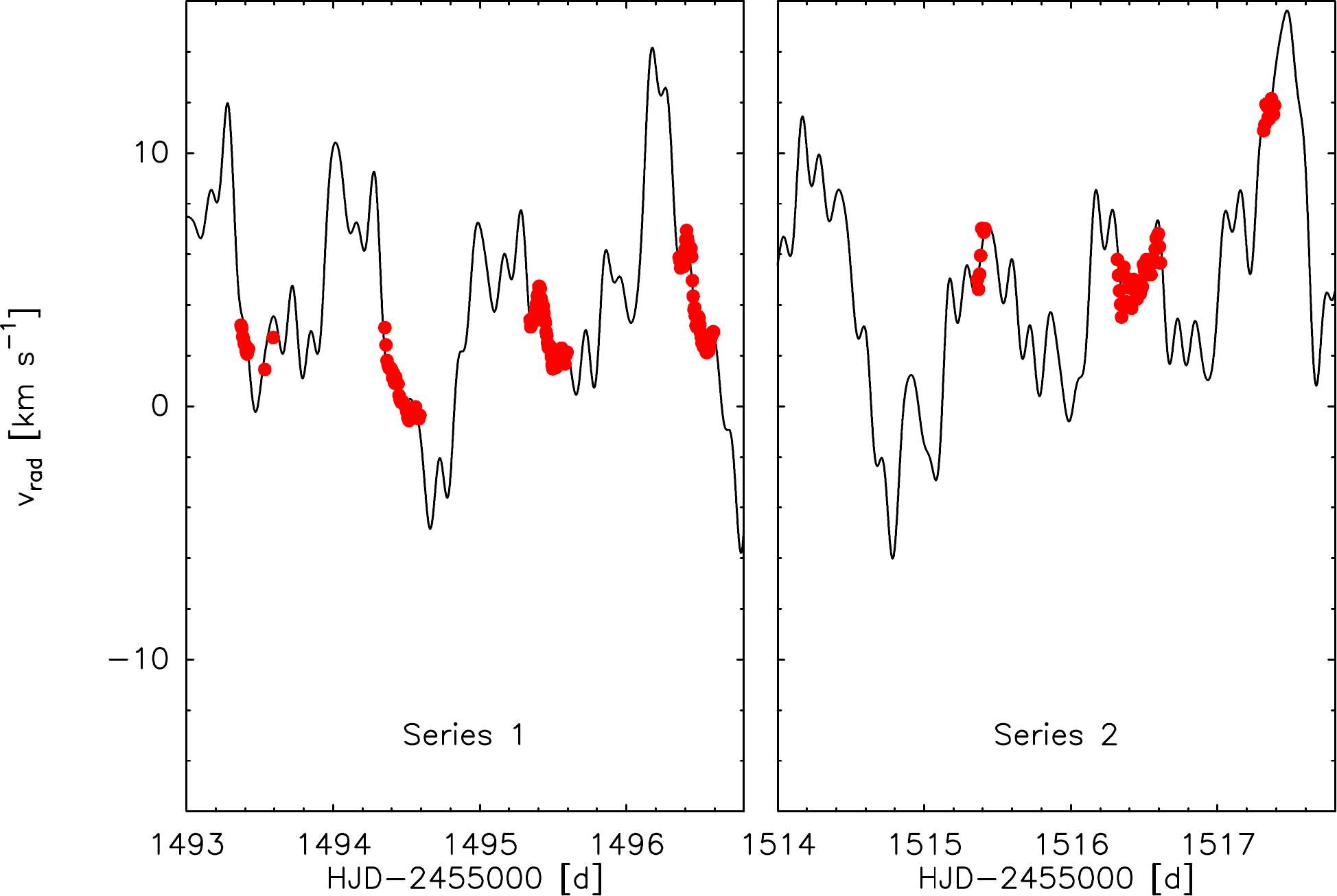}
      \caption{Fit (solid line) to the radial velocity variations (dots) of
         55\,Cyg in the total period 2013 July 19 to August 12 (top) and 
         zoomed in for the two marked series (bottom).}
         \label{fig:fit_complete}
  \end{figure}

\subsubsection{Frequency analysis}
\label{frequency_analysis}
The results from the moment analysis suggest pulsations as the most plausible
cause for the observed line profile variability in He, as well as in the
metal lines. The clear next step is to analyze the radial velocity curve to
search for periods. We performed a frequency analysis using a similar strategy
as the one described by \citetads{Saio2006}.
We first computed a
Lomb-Scargle periodogram \citepads{1982ApJ...263..835S}
for frequency identification and amplitude estimation, which is displayed
in the top panel of Fig.\,\ref{fig:ampli}. The periodogram finds no single
dominant period. Instead, it seems that multiple significant peaks are present.
Such a behavior is known for the $\alpha$ Cygni variables
\citepads{1976ApJ...206..499L}, for instance, and has also been reported for the periodograms of
other BSGs \citepads[e.g.,][]{Saio2006}, and we therefore used a fitting function
consisting of a series of sinusoids. To refine the parameters for the
identified frequency, we applied a nonlinear least-squares fit. A sine
curve was fit to the data and subtracted. Then we searched for the next frequency
within the residuals. In each subsequent cycle, the sum of all previously
identified sine curves was subtracted from the original data. The procedure was
stopped as soon as the S/N of the next frequency was too low.

We define the noise level $\sigma_{\rm res}$ as the mean amplitude of the
residual in a frequency interval around the identified frequency. The noise
level in the frequency range 0 to 10 cycles day$^{-1}$ is 0.13\,km\,s$^{-1}$.
Frequencies with an amplitude $A > 4\sigma_{\rm res}$ correspond to a
confidence level of 99.9\%, and are considered as clear detections. Those with
$A > 3.6\sigma_{\rm res}$ correspond to a confidence level of 95\%.
In total, we identified 19 frequencies. Only the last one has an amplitude
of $3.7\sigma_{\rm res}$, and should thus be considered as less accurate.
All other detections have amplitudes $A > 4\sigma_{\rm res}$.

\begin{figure}
   \centering   
   \includegraphics[width=\hsize]{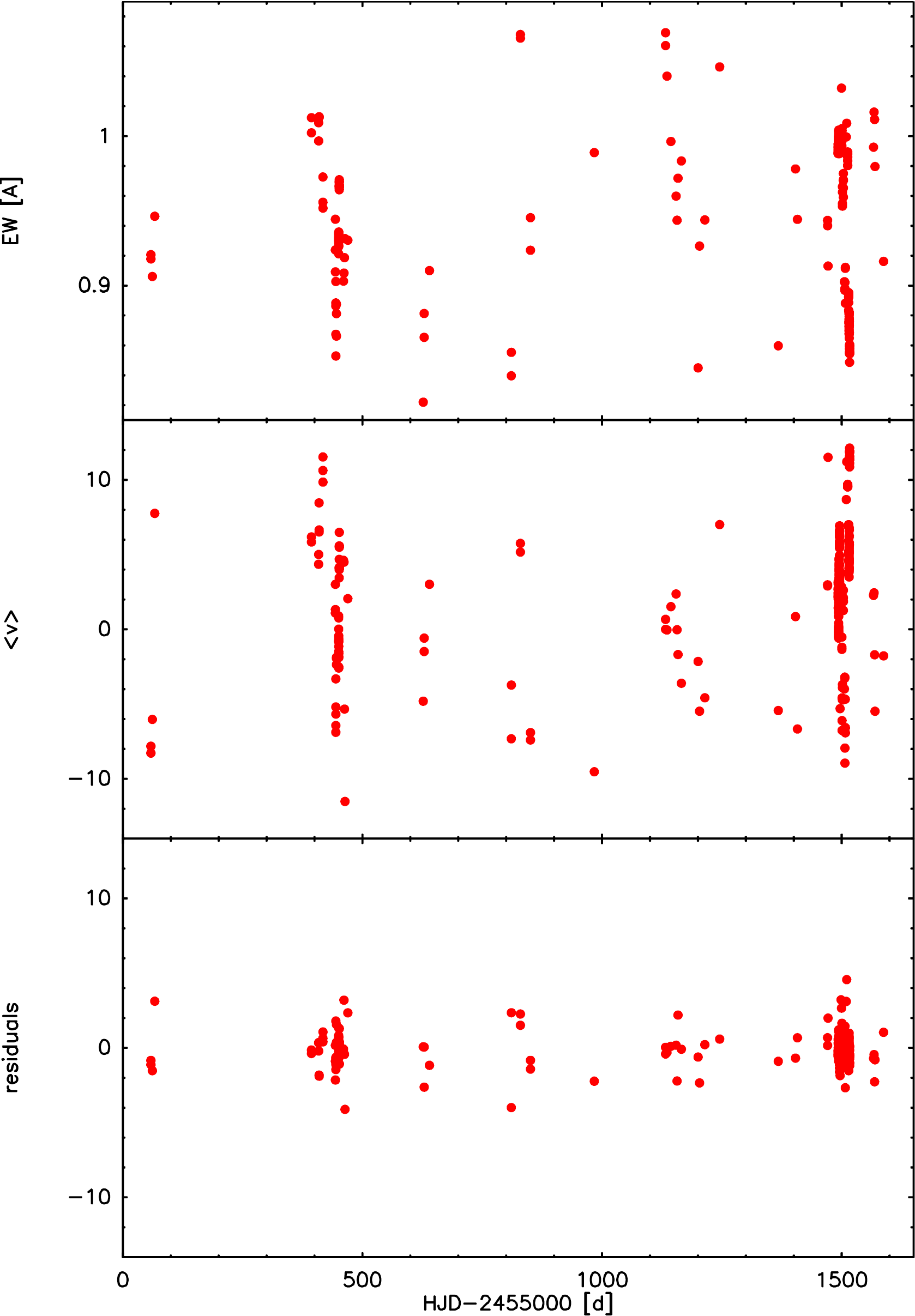}
      \caption{Variation in EW (top) of the \ion{He}{i}\,$\lambda$\,6678\,\AA\,
        line over the total observing period, the radial velocity
        (middle) and the residuals of the radial velocity (bottom) after
        pre-whitening with the superposition of 19 sinusoids.}
      \label{fig:he_resid}
\end{figure}

\begin{figure}
   \centering   
   \includegraphics[width=\hsize]{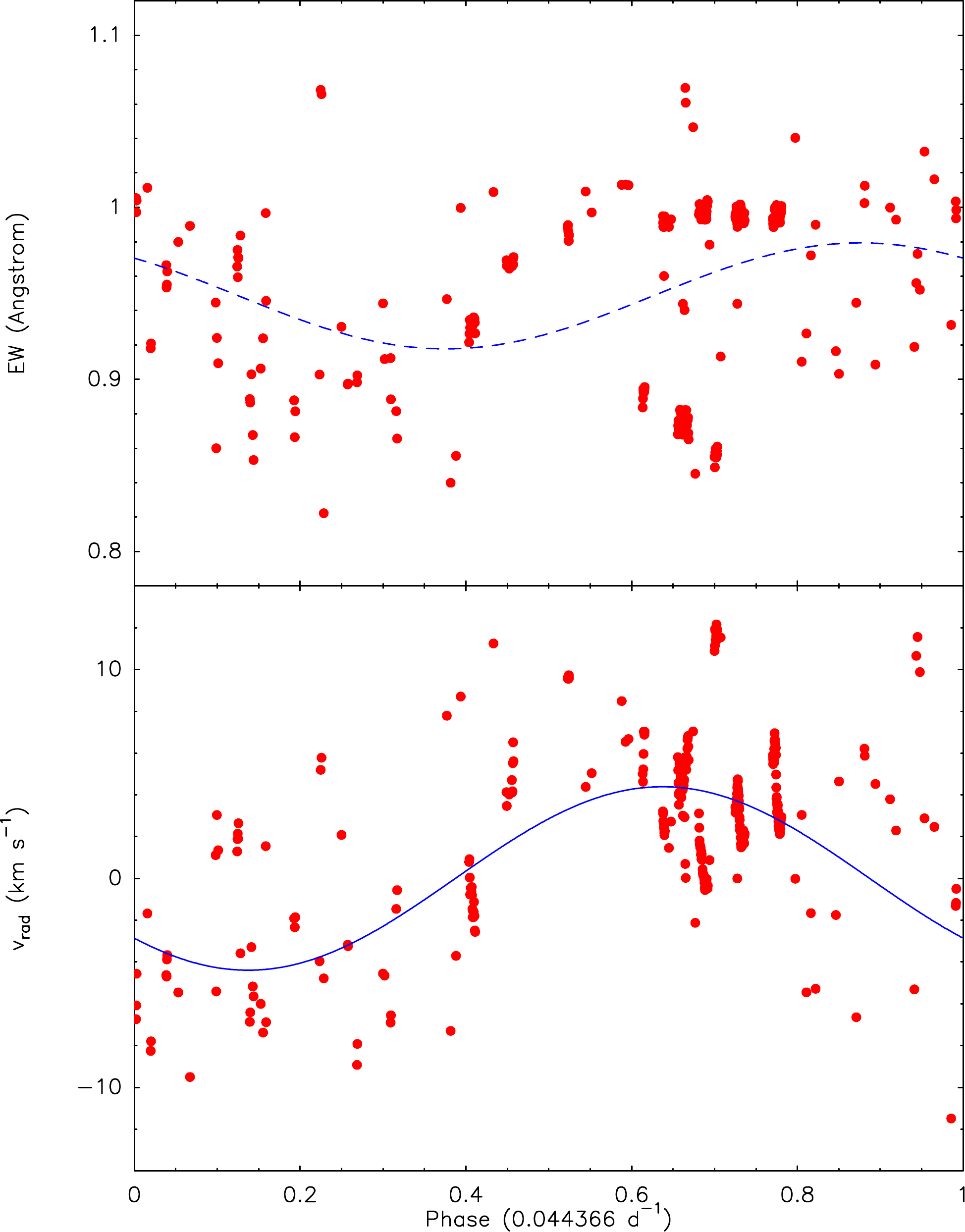}
      \caption{Radial velocity and EW of the \ion{H}{i}\,$\lambda$6678 line
       phased to the longest period (22.5\,d). The corresponding sine curve
       with the parameters from Table\,\ref{tab:frequencies} is included in the
       phase diagram of the radial velocity, and a tentative fit to the EW, which reveals the same period but is phase-shifted.}
      \label{fig:phase}
\end{figure}

With all frequencies identified, we again performed a least-squares fit
using the parameters of all periods as first-guess values. The final refined
parameters of each sine curve with their errors 
are listed in Table\,\ref{tab:frequencies}. The Lomb-Scargle
periodogram of the data, pre-whitened with the 19 identified
periodicities, is shown in the bottom panel of Fig.\,\ref{fig:ampli}.

Next, we computed the sum of the 19 individual sinusoids as a function of time
(i.e., HJD) and compared it to the observations. The synthetic radial velocity
curve looks rather chaotic.
For clarity and better visibility, we plot the synthetic curve limited to
a period of 25 days in the top panel of Fig.\,\ref{fig:fit_complete} and
overplot the values obtained from the observations. In this plot, we mark
two series, which correspond to the time interval where we obtained
time series of observations of four and three consecutive nights. The comparison
between the model and the observations of these series is shown, zoomed in, in
the bottom panels of Fig.\,\ref{fig:fit_complete}.

In addition to the radial velocity of the \ion{He}{i}\,$\lambda$6678 line, we also 
computed the line EW. Its value over the complete four-year observing 
period is displayed in the top panel of Fig.\,\ref{fig:he_resid}. It is 
found to be quite variable as well, which is an indication for temperature
variations in the stellar atmosphere \citepads[e.g.,][]{2002A&A...385..572D}. 
We phased the data (both EW and first moment) to the longest period of 
22.5\,days identified from the radial velocity curve.
This phase diagram is shown in Fig.\,\ref{fig:phase}. The radial velocity
clearly displays a sinusoidal behavior (bottom panel, with the sine curve of 
the identified period overplotted). We reveal variability with the same 
period in EW (top panel), but with a phase shift of 
$94\degr$. Considering that this period could belong to a radial (strange) 
mode, a phase shift in EW of $90\degr$ would imply an 
adiabatic pulsation mode, while slight deviations are usually interpreted as 
non-adiabatic effects. Moreover, the large scatter in the
phased EW can be explained by the multiperiodicity \citepads{2002A&A...385..572D}.

\section{Discussion}
\label{sec:disc}

We interpret the spectroscopic variability observed in the blue supergiant star 
55\,Cyg in terms of photospheric oscillation modes and a time-variable 
radiation-driven wind. We discuss the reliability of the identified frequencies 
and the variation of wind parameter models. We also discuss the possibility that 
photospheric and wind variations might be related.

\subsection{Reliability of the frequencies}

The agreement between the synthetic radial velocity curve and the observations 
seems to be quite good (see Fig.\,\ref{fig:fit_complete}), but not all 
data fall on the model curve. This is also visible from the final residuals, 
which still display some scatter. These residuals are shown for the full 
observing period in the bottom panel of Fig.\,\ref{fig:he_resid}, in comparison 
to the original observed radial velocity variation (middle panel).

There are three possible reasons why we did not obtain better fits to the radial
velocity curve. First, the number of observations is not sufficient to properly
determine the stellar radial velocity component. Hence, the data might not be
exactly symmetric around zero, influencing the determination of the frequencies
and amplitudes of the periods. Second, not all data points have the same high
quality, and there are a number of rather large gaps in the observations. Both
lead to an increase in the noise level of the amplitude spectrum. Hence, from
ground-based spectroscopy we identify only 19 frequencies, while (many) more
might be present. For comparison, 48 frequencies were identified in the
supergiant \object{HD\,163\,899} by \citetads{Saio2006}, using space-based 
photometry. Third, nothing is known about the stability of the periodic 
variations in 55\,Cyg. As the observing period stretches over more than four 
years, some of the periodicities might have changed or completely disappeared,
while others might have appeared. Such a behavior of flipping periodicities was 
reported by \citetads{2008A&A...487..211M} for the BSG star 
{\object{HD\,199\,478}. Furthermore, \citetads{Aerts2010} recognized a sudden 
change in the amplitude of the pulsation mode in the BSG star HD\,50\,064.

Consequently, precise determinations of currently present periodicities in  
55\,Cyg must be obtained from a better dataset, ideally from 
continuous spectroscopic observations over a long time interval, combined with 
a continuous photometric light curve. This is a non-trivial task,
however.

\subsection{Stellar rotation}
\label{dis-rotation}
\begin{figure}
   \centering  
   \includegraphics[width=\hsize,angle=0]{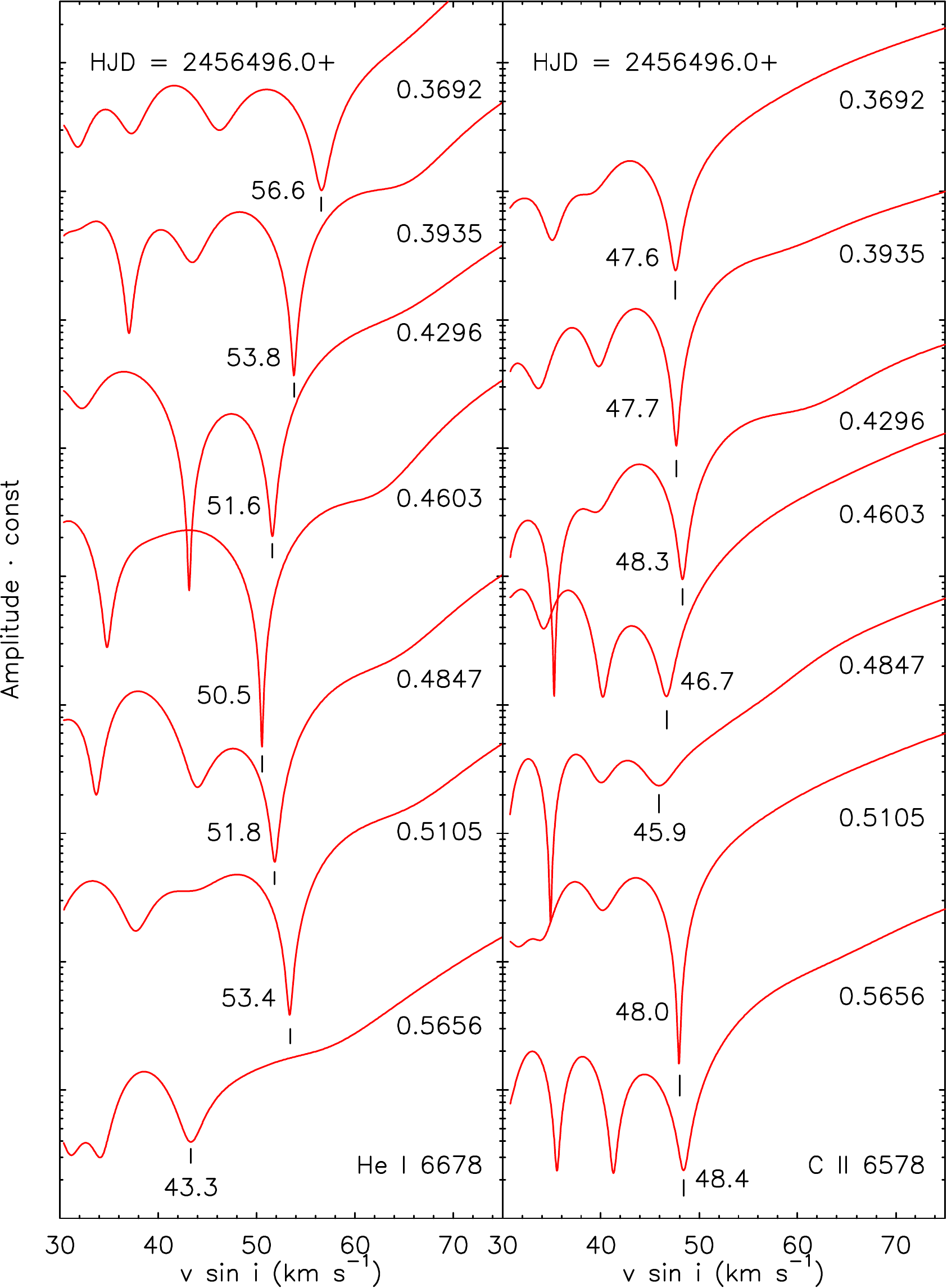}
   \includegraphics[width=\hsize,angle=0]{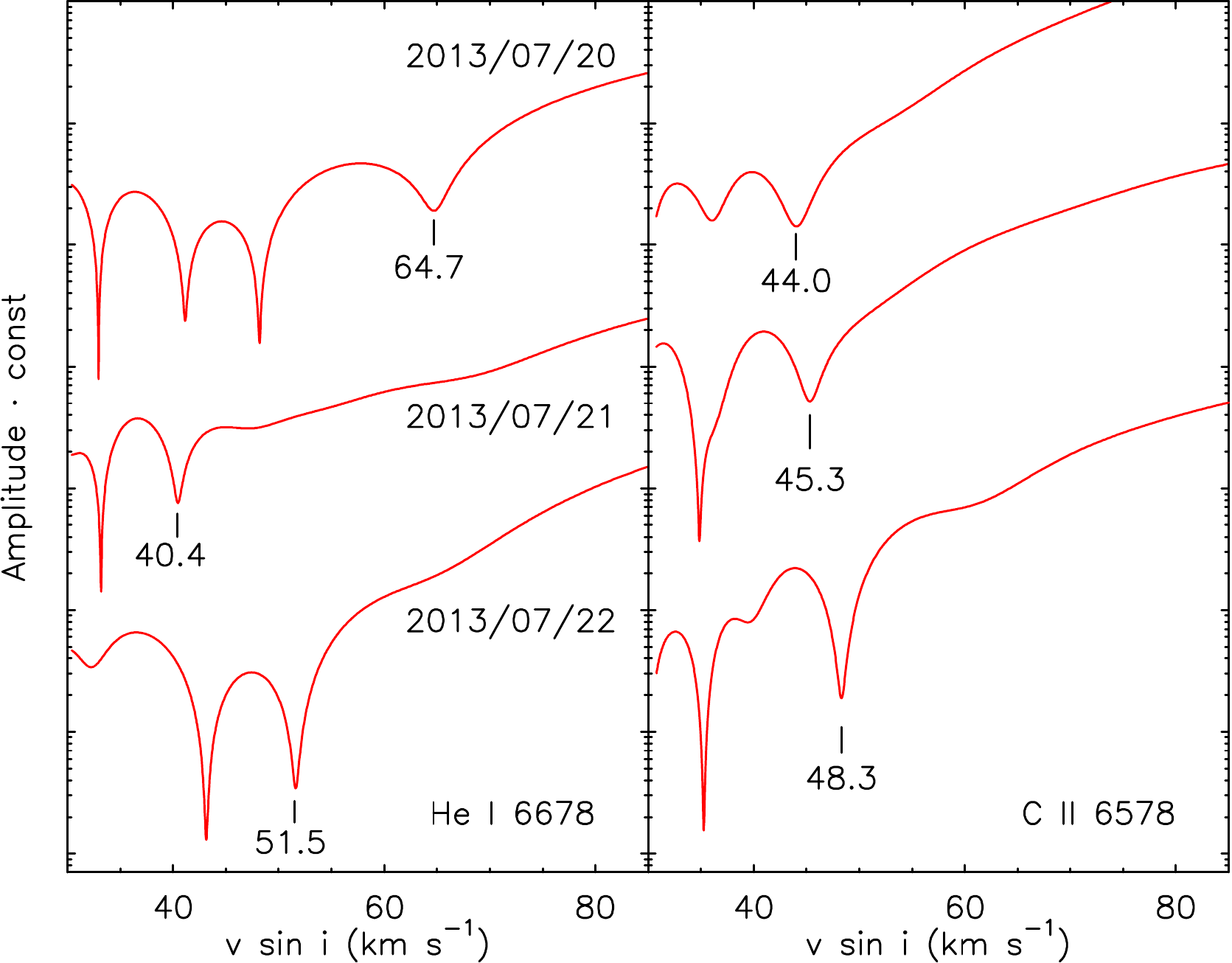}
      \caption{Top: Selection of values of $v\sin i$ obtained with the
        FT method for \ion{He}{i}\,$\lambda$6678\,\AA\, (left) and 
        \ion{C}{ii}\,$\lambda$6578\,\AA\, (right) within the
        night of 2013 July 22. Only results with a clear first minimum
        are shown. Bottom: Range in $v\sin i$ values obtained within three
        consecutive nights.}
         \label{fig:vsini}
\end{figure}

The presence of multiple periodicities with significant velocity amplitudes in 
the photospheric lines implies that their profiles are affected in two ways.
First, the lines can be expected to be broadened (referred to as 
macroturbulence) in excess of 
the typical broadening mechanisms acting in stellar atmospheres, including 
stellar rotation. And second, the profile shape is
usually not symmetric anymore. The latter is confirmed from our moment
analysis (see Fig.\,\ref{fig:he_mom}). 
However, the amounts of both rotation and macroturbulence are not known 
initially. The determination of proper stellar rotation rates is particularly 
important because rotation plays a major role in the evolution and mass-loss 
behavior of massive stars \citepads[see, e.g.,][]{2000ARA&A..38..143M, 
2005A&A...429..581M, 2012A&A...537A.146E, 2014A&A...562A.135S}.

Investigations based on different methods, such as the Fourier transformation 
(FT) or the goodness-of-fit (GOF), are usually applied with the aim to 
determine proper $v\sin i$ values and the contribution of macroturbulence.
However, a degeneracy exists, which allows fitting observed line profiles 
with a diversity of different combinations of $v\sin i$ and $v_{\rm macro}$
\citepads[e.g.,][]{Ryans2002}. This degeneracy was also found for 55\,Cyg by 
\citetads{Jurkic2011} from modeling silicon lines in IUE spectra. Furthermore, 
the major requirement for a reliable result from the FT method is a symmetric 
signal. Application to asymmetric line profiles, as is typically expected for 
pulsating stars, immediately delivers false results. Studies of
the effect of pulsational activity in stellar atmospheres on resulting
line profiles revealed that the collective effect of multiple periodicities,
for example, in the form of low-amplitude non-radial gravity modes, can cause
substantial line broadening, which might be identified with macroturbulence
\citepads{Aerts2009, 2010ApJ...720L.174S}. These investigations also revealed 
that the FT method fails as soon as pulsations play a non-negligible role
in the line width. Moreover, \citetads{2014A&A...569A.118A} showed that 
results from the FT of line profiles of known pulsators vary substantially over 
the pulsation cycle.

Even though our observations have extremely good S/N, which is
considered as a major prerequisite for reliable results from the FT
\citepads{Simon-Diaz2007}, it is not possible to determine consistent 
$v\sin i$ values for 55\,Cyg. This is shown in 
Fig.\,\ref{fig:vsini}, where we plot the results from the FT analysis of 
\ion{He}{i}\,$\lambda$6678\,\AA\, and \ion{C}{ii}\,$\lambda$6578\,\AA\,
during different times within one night (top panel) and extreme values 
obtained during three consecutive nights (bottom panel). For many line
profiles (for both \ion{He}{i} and \ion{C}{ii}) 
the FT of our data deliver no clear minimum 
and, hence, no value for $v\sin i$ could be found. 

From the profiles with clear zero points in their FT,
\ion{C}{ii} suggests $v\sin i$ values between 44\,km\,s$^{-1}$ and 
48\,km\,s$^{-1}$ (which resemble the rotational values of 40\,km\,s$^{-1}$ and 
45\,km\,s$^{-1}$  obtained for the \ion{Si}{ii} lines). The values obtained from the \ion{He}{i} line are in 
most cases significantly higher and with a higher spread in values, for example,
from about 40\,km\,s$^{-1}$ to about 65\,km\,s$^{-1}$. While a shift in 
the position of the first zero point and the appearance of saddle points 
and local minima in the FT can be expected in the case of pulsations
\citepads{2014A&A...569A.118A}, the discrepancy in the values obtained from
the two investigated lines is difficult to understand as the FT method 
usually results in very similar $v\sin i$ values when applied to  He
and metal lines \citepads{Simon-Diaz2007}. While the variation in values
obtained from \ion{C}{ii} is small, it may still imply that macroturbulent
line broadening (and hence the underlying pulsations) influences the position
of the zero points. In that context, strongly
varying $v\sin i$ values within a single night were obtained from the 
FT method applied to the \ion{Si}{ii}\,$\lambda$\,6347 line of the 
BSG HD\,202\,850 by \citetads{Tomic2015}.

In addition to the seemingly higher $v\sin i$ value resulting from the FT, the 
\ion{He}{i}\,$\lambda$6678\,\AA \ line also appears to be special in other aspects. 
For instance, the measured radial velocity is higher than in the 
\ion{C}{ii}\,$\lambda$6578\,\AA \ line (see Fig.\,\ref{fig:metall}).
Moreover, \ion{He}{i}\,$\lambda$6678\,\AA \ is the only line in our sample 
(apart from H$\alpha$) that requires substantially higher values for both the
micro- and macroturbulent velocities than all other modeled photospheric
lines. A possible explanation could be that the \ion{C}{ii}\,$\lambda$6578\,\AA \ 
and the \ion{He}{i}\,$\lambda$6678\,\AA \ lines are formed in different 
regions in the atmosphere with \ion{C}{ii} originating in deeper layers, in 
agreement with the \ion{C}{ii} line being much weaker than the \ion{He}{i} 
line (see Fig.\,\ref{fig:spectrum}). 
The line-forming region for \ion{He}{i} might be substantially larger
and may be extending into the base of the wind, where higher turbulent 
velocities (i.e., higher than in the photosphere) might
be expected due to instabilities, possibly resulting from coupling of
the pulsation modes.

Although the FT analysis of the line profiles has severe problems in the
presence of strong macroturbulence and possible wind contamination of the 
profiles, the range of values obtained from both lines is on the same order
as the value of $55\pm 5$ km\,s$^{-1}$ found from the line profile fits
(see Sect.\,\ref{vel}).
However, as long as the pulsation modes cannot be identified, the individual
contributions from the two different, but physically very important mechanisms
acting on the line profiles, that is, rotation and pulsations, remain uncertain.

\subsection{Connection between pulsations and mass-loss}

Our analysis revealed multiple periods, which 
can be interpreted as stellar pulsations. Hence, the question arises whether these 
pulsations might be suitable to trigger time-dependent enhanced mass loss, 
a question that was also addressed by \citetads{Lefever2007} for a large sample of BSGs with
no firm answer. 

Two frequencies in the set of frequencies identified in Sect.\,\ref{frequency_analysis}, f$_3$ (4.2 days) and f$_{11}$ (5.3 days), are close to the 
spectroscopic period of 4--5 days reported by \citetads{1975A&A....45..343G} 
and the periodicity of 4.88 days found in the photometric light curve by  
\citetads{2002MNRAS.331...45K}. Twelve frequencies correspond to  oscillations 
with periods spreading from  $\sim 2.7$\,hours to  $\sim\,$24 hours. The 
frequencies f$_5$ and f$_6$ might be harmonics of  f$_{15}$ and  
f$_{16}$, respectively.  Meanwhile, the frequencies associated with periods of 
about one day (f$_2$ and f$_4$) might be related to the daily variations 
observed in the intensity of the H$\alpha$ line because their similar 
frequencies cause beats.

We would like to stress that the daily changes in line profile shape and 
intensity seen in our data should not be interpreted as a real change in 
mass loss. Each of our spectra has been modeled 
independently, considering a spherical smooth wind with a $\beta$ velocity
law. However, any density enhancement at the base of the wind due to some
mass ejection (triggered by pulsation or any other instability) will need
time to travel through the H$\alpha$ line-forming region and hence 
contribute to the H$\alpha$ line profile in more than just one night. The
night-to-night variations that we see in the data can therefore not be 
regarded as real changes in the wind parameters (mass loss and terminal 
velocity) but should rather be interpreted as rapid changes in the wind 
density, that is, wind inhomogeneities, traveling through the wind and 
influencing the H$\alpha$ line profile.

It is also worth mentioning that there were three occasions at which
the H$\alpha$ line completely (or almost completely) disappeared from the 
spectrum, indicating that in these particular cases the wind emission was 
reduced so that it compensated by coincidence for the photospheric absorption line. 
The first time noted was on 2010 July 2-4 by \citetads{Maharramov2013}, and in 
our own data we recognized it on 2012 July 24
and in 2013, around July 26. 
Interestingly, after the re-appearance of the H$\alpha$ in emission, its 
profile does not recover the shape seen before the cancellation. 
Instead, a strong, pure emission evolves during 2013 August 10-12, which might 
be explained as due to an enhanced amount of material appearing in the 
H$\alpha$ line-forming region. Furthermore, inspection of the behavior of the 
\ion{He}{i}\,$\lambda$6678 line in the same period reveals a strong increase in 
the radial velocity (middle panel of Fig.\,\ref{fig:he_mom}) with a 
simultaneous drop in EW (top panel of Fig.\,\ref{fig:he_mom}).

If the occasions of annihilation of emission and absorption components of the H$\alpha$ line 
were periodic, the possible period would be on the order of $ \text{about  one}$\,year. 
We may speculate that the
wind conditions of 55\,Cyg may have changed some years ago, because no earlier 
observational work on 55\,Cyg mentions phases in which the wind emission replenishes the 
photospheric absorption line. 

On the other hand, according to the stellar parameters derived in this work, 
$\log\,L/L_{\odot} = 5.57$, and the current stellar mass of 34\,M$_{\odot}$, 
we find a ratio for $L/M\,\sim\,1.1\times 10^{4}$\,L$_{\odot}$/M$_{\odot}$ 
, which is high enough for the occurrence of radial strange modes 
\citepads{Gautschy1990,Glatzel1994,Saio1998}. 
The stellar properties of 55\,Cyg resemble those of HD\,50\,064, for which 
\citetads{Aerts2010} argued that the star is building up a circumstellar 
envelope, as seen in the variable Balmer lines, while 
pulsating in a radial strange mode. A similar scenario might also hold for 55\,Cyg. 
The first frequency, f$_1$ (22.5 days), has the proper length to be 
interpreted as radial strange mode as predicted from theory \citepads{Saio2013}.
This period might tentatively be connected to the mid-term variation observed 
in the H$\alpha$ line and the mass-loss rates. Considering the sequence of 
observations obtained in October 2013 in Arizona, it seems that the mass-loss 
rate may keep more or less a constant value over a time interval of roughly a 
week, followed by a jump to a lower mass-loss regime. This second regime might 
have lasted seven or perhaps more days (neglecting the sudden jump in the mass-loss 
observed on October, 30). Another period of at least five days of constant 
behavior can be recognized in July 2013. Whether this (quasi-)periodic
variability in mass-loss might be recovered in other seasons cannot be asserted 
because of the often large gaps in observations.

Additional support for a possible connection between pulsations and enhanced mass loss 
is provided by the variable line shapes of H$\alpha$ and by the faint shallow symmetric 
wings extending to $\pm 1200$\,km\,s$^{-1}$ (described in Sect.\,\ref{sec:obs}).
The velocity of these wings is definitely higher than the wind terminal velocities
determined for 55\,Cyg, so that they are not formed by the current wind material.
Instead, we interpret these wings as due to incoherent electron scattering 
in the outer layers of the star's extended envelope above the line formation
region.
As the wings were not reported by other observers \citepads[with one exception, 
see][]{1982ApJS...48..399E}, but are rather stable during 
the past four years, the pulsational activity leading to the formation of an
optically thick, expanding atmosphere might have started or increased within 
the past few years.

\subsection{Evolutionary state of  55\,Cyg}

According to the derived effective temperature and luminosity, we might connect the star with an evolutionary track of 40 M$_\odot$. The assignment of a proper evolutionary state of the star is hampered by the fact that massive stars can cross the temperature range of BSGs more than once during their evolution. 
The location of a star in the HRD alone is thus not sufficient, and additional criteria are needed for an unambiguous evolutionary
state determination. One such criterion is provided by the recent calculations of the occurrence of possible pulsation modes during the post-main sequence evolution of massive stars by \citetads{Saio2013}. Their investigations
revealed that during the post-RSG evolution, BSGs can pulsate in many more modes and, in particular, in radial strange modes, which is not possible while
the star is on its first redward evolution. Moreover, the observations of pulsation properties with very many excited modes are possible only after the star has experienced a strong mass-loss during the RSG phase.

The numerous periodicities identified in 55\,Cyg display a mixture of 
p- and g-modes.
According to computations of pulsation modes in BSGs, for rotating
stellar models in the range of initial masses of 20-25\,M$_{\odot}$yr$^{-1}$ 
\citepads{Saio2006, Saio2013}, the co-existence of both types of modes 
in the pre-RSG stage is found for stars with temperatures $T_{\rm{eff}} \ga 
20\,000$\,K, while cooler stars are expected to display very few pulsations.
Even though the temperature of 55\,Cyg, $T_{\rm{eff}}$ = 19\,000 K, is slightly
lower, its periodicities are too numerous to fit in a pre-RSG scenario.
Consequently, a post-RSG stage is more plausible mainly because of the 
multiple pulsation modes (with periods $\la 30$\,days) that we find in this star. 
Furthermore, the classification of 55\,Cyg as a post-RSG is supported by 
the presence of the bow shock feature detected by the Wide-field Infrared 
Survey Explorer \citepads[WISE,][]{Wright2010} in band 4 at 24 $\mu$m 
(see Fig. \ref{fig-wise}), which formed in the past by the interaction of a strong 
wind phase with the ISM.

But this post-RSG scenario does not seem to fit the low He surface 
abundance we derived from our line profile modeling. This apparent 
contradiction was also found in at least two other BSGs \citepads[\object{Rigel}
and \object{Deneb}, see][]{Saio2013}. To solve this problem, \citetads{Georgy2014} 
recently showed that new stellar evolution calculations, using the Ledoux 
criterion for convection, reconcile the expected pulsational properties of a 
post-RSG phase with lower surface He chemical abundance (values that were 
previously expected only during the pre-RSG phase assuming the Schwarzschild 
criterion for convection).  

\begin{figure}
\setlength\fboxsep{0pt}
\setlength\fboxrule{0.5pt}
\begin{center}  
\includegraphics[width=0.98\linewidth]{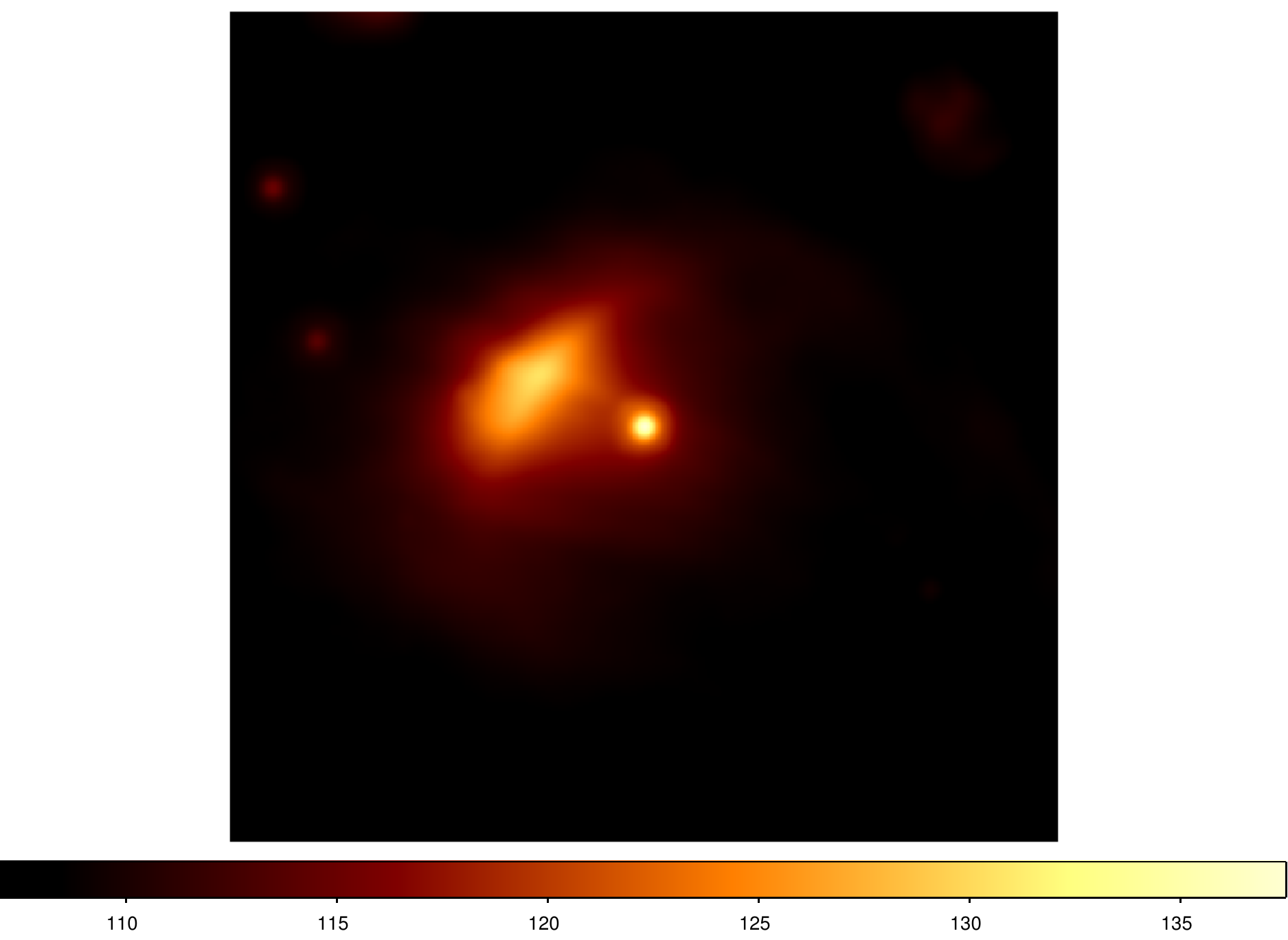}
\caption{Bow shock features detected by WISE in band 4 at 24 $\mu$m. The size 
of the image is $10\arcmin\times 10\arcmin$. \label{fig-wise}}
\end{center}
\end{figure}

\section{Conclusions}
\label{sec:concl}

55\,Cyg shows both long- and short-term variability in its spectroscopic and 
photometric data. WISE IR images reveal a large bow shock structure originating 
from the interaction of a strong mass-loss event with the ISM. This strong wind 
phase is expected to occur during evolution through the RSG stage.

To understand the origin of the variability observed in the photospheric lines,
we performed a moment analysis based on almost 400 optical spectra spread over$\text{ about five}$ years and identified 19 periods (although many more might be present). 
Twelve of these periods spread from $\sim 2.7$\,hours to $\sim\,$24 hours and 
might belong to p-modes, and 6 are between 1 and 6 days long and could represent 
g-modes. The longest period of 22.5 days could be related to the theoretically
predicted strange mode pulsations. Support for this interpretation comes from 
the value of the $L/M$ ratio, which is of the correct order of magnitude for 
strange mode excitation. The existence of many short pulsation modes might 
influence the line profile of the photospheric lines, and might, hence, be 
responsible for the large uncertainties in the derived rotational velocity of 
the star and the variety of macroturbulent velocities measured in different lines.

We modeled the H, \ion{He}{i}, \ion{Si}{ii,} and \ion{Si}{iii} lines using 
the NLTE atmosphere code FASTWIND and derived stellar and wind parameters. We
found that 
the star undergoes episodes in which the determined mass-loss can vary by a 
factor of 1.7--2 on timescales of about two to three $ $ weeks. We also detected a 
long-term period in which the star exhibits no H emission lines. In the 
H$\alpha$ line profile, we detected strong night-to-night intensity and profile 
shape variability, indicating rapid changes in the wind that might be 
interpreted as local material 
enhancements, caused by short-term mass ejection events traveling through the H$\alpha$ line-forming region.
Multiple periods in both the wind and photospheric material might
suggest a possible link between pulsational activity and enhanced mass-loss 
in  55\,Cyg.
Moreover, combining the high derived spectroscopic mass of M$_\star = 
34$\,M$_\odot$, the many excited pulsation modes, the evidence of a 
huge past mass-loss phase, and the presence of CNO processed material at the 
surface, 
55\,Cyg is most probably in a post-RSG phase, crossing the HRD toward the blue 
region. As such, the star conforms to characteristics resembling classical 
$\alpha$\,Cygni variables \citepads{Georgy2014,Saio2013}.

\begin{acknowledgements}
We thank the technical staff at the Ond\v{r}ejov Observatory for the support during the observation periods, and Ad\'{e}la Kawka for taking some of the spectra.
We acknowledge Joachim Puls for allowing us to use the FASTWIND code, for 
his help and advices with the code, and for his comments and suggestions on the 
paper draft. This research made use of the NASA Astrophysics Data System (ADS) and of the SIMBAD database, operated at CDS, Strasbourg, France. M.K., D.H.N.,
and S.T. acknowledge financial support from GA\,\v{C}R (grant numbers 209/11/1198 and 14-21373S). The Astronomical Institute Ond\v{r}ejov is
supported by the project RVO:67985815.
L.C. acknowledges financial support from the Agencia de Promoci\'on Cient\'{\i}fica  y Tecnol\'ogica (Pr\'estamo BID PICT 2011/0885), CONICET (PIP 0300), and the Universidad Nacional de La Plata (Programa de Incentivos G11/109), Argentina.
Financial support for International Cooperation of
the Czech Republic (M\v{S}MT, 7AMB14AR017) and Argentina (Mincyt-Meys ARC/13/12 and CONICET 14/003) is acknowledged. A.A. acknowledges financial support from 
the Estonian Ministry of Education and Research. E.N. would like to acknowledge 
the NCN grant  No. 2014/13/B/ST9/00902. M.C. acknowledges, with thanks, the 
support of FONDECYT project 1130173 and Centro de Astrof\'isica de 
Valpara\'iso. The research leading to these results has also received funding 
from the European Research Council under the European Community's Seventh 
Framework Programme (FP7/2007--2013)/ERC grant agreement N$^{\underline{\mathrm
o}}$\,227224 ({\sc prosperity}) and from the Czech Ministry of Education
(M\v{S}MT, project LG14013). Observational work of Poznan Spectroscopic 
Telescope 2 was supported by Polish NCN grant UMO-2011-01/D/ST9/00427.
\end{acknowledgements}

\bibliographystyle{aa} 
\bibliography{citas} 

\end{document}